\documentclass[fleqn,usenatbib]{mnras}
\usepackage{graphicx}
\usepackage{natbib}
\usepackage{amsmath}
\usepackage{amssymb}
\usepackage{verbatim}
\usepackage{url}
\usepackage{threeparttable}
\usepackage[T1]{fontenc}
\usepackage{ae,aecompl}

\newcommand       \pd		[2]       {\frac{\partial #1}{\partial #2}}
\newcommand       \sub       [1]       {_\textrm{#1}}

\title[Balancing the Budget in Pebble Accretion]{A Balanced Budget View on Forming Giant Planets by Pebble Accretion}
\author[Lin, Lee, \& Chiang]{Jonathan W. Lin$^{1,2}$\thanks{Contact e-mail: \href{mailto:jon.880@berkeley.edu}{jon.880@berkeley.edu}}, Eve J. Lee$^{1,3}$\thanks{\href{mailto:evelee@caltech.edu}{evelee@caltech.edu}}, Eugene Chiang$^{1,4}$\thanks{\href{mailto:echiang@astro.berkeley.edu}{echiang@astro.berkeley.edu}}\\
$^{1}$Department of Astronomy, University of California Berkeley, Berkeley, CA 94720-3411, USA\\
$^{2}$Department of Engineering Science, University of California Berkeley, Berkeley, CA 94720-1702, USA\\
$^{3}$TAPIR, Walter Burke Institute for Theoretical Physics, Mailcode 350-17, Caltech, Pasadena, CA 91125, USA\\
$^{4}$Department of Earth and Planetary Science, University of California Berkeley, Berkeley, CA 94720-4767, USA
}

\date{}
\pagerange{\pageref{firstpage}--\pageref{lastpage}} \pubyear{2018}
\begin{document}
\maketitle
\label{firstpage}

\begin{abstract}
Pebble accretion refers to the assembly of rocky planet cores from
particles whose velocity dispersions are damped by drag from
circumstellar disc gas. Accretion cross-sections can approach
maximal Hill-sphere scales for particles whose Stokes numbers
approach unity. While fast, pebble accretion is also lossy. Gas drag
brings pebbles to protocores but also sweeps them past; those
particles with the largest accretion cross-sections also have the
fastest radial drift speeds and are the most easily drained out of
discs. We present a global model of planet formation by pebble
accretion that keeps track of the disc's mass budget. Cores, each initialized
with a lunar mass, grow from discs whose finite stores of mm--cm
sized pebbles drift inward across all radii in viscously accreting
gas. For every $1 M_\oplus$ netted by a core, 
at least 10$M_\oplus$ and possibly much more
are lost to radial drift. Core growth rates are typically
exponentially sensitive to particle Stokes number, turbulent Mach
number, and solid surface density. This exponential sensitivity, when
combined with disc migration, tends
to generate binary outcomes from 0.1--30 AU:
either sub-Earth cores remain sub-Earth, or
explode into Jupiters, with the latter migrating
inward to varying degrees. 
When Jupiter-breeding cores assemble from mm--cm sized pebbles, 
they do so in discs where such particles drain out in
$\sim$10$^5$ yr or less;
such fast-draining discs do not fit mm-wave observations.
\end{abstract}

\begin{keywords}
planets and satellites: formation, protoplanetary discs
\end{keywords}

\begin{figure*}
 \includegraphics[width=16cm]{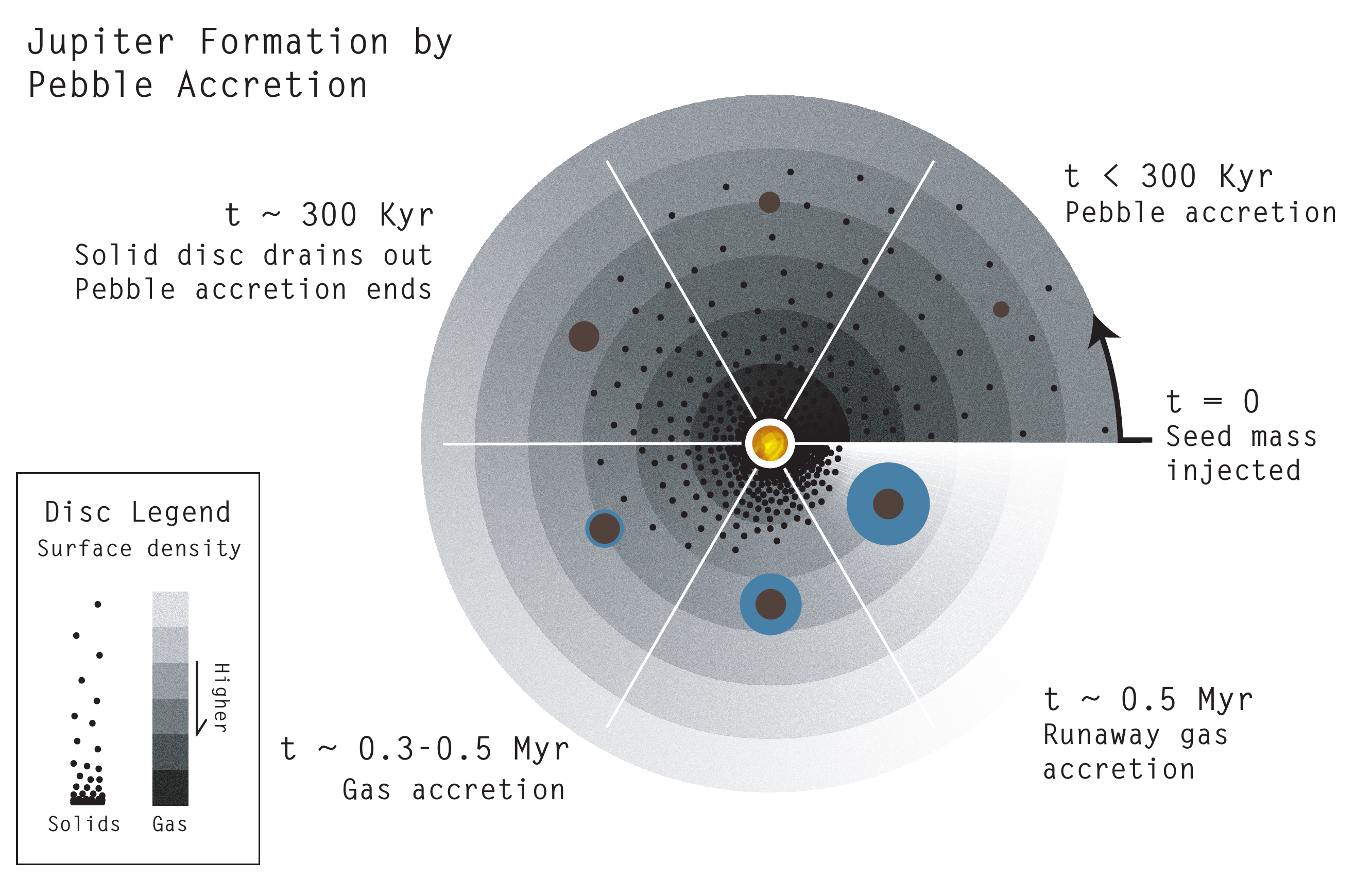}
\caption{
An example formation pathway for Jupiters, abstracted from the
quantitative model shown in Figure \ref{fig:wj}.
Each sector represents a different temporal snapshot of the disc and planet,
to be read counterclockwise starting from 3 o'clock.
A seed core (brown) is placed in a model gas disc (grey) and begins accreting
solid particles (''pebbles''; black). At the same time that pebbles
feed the core,
they drift inward by aerodynamic drag exerted by the ambient gas disc. The
core stops growing once the outer edge of the solid disc sweeps by.
When core assembly ends and heating from pebble accretion subsides, a
gas envelope (blue) grows by Kelvin-Helmholtz cooling and contraction.
Eventually, the planet's gas mass becomes comparable to its core mass,
the atmosphere's cooling time shortens catastrophically because of
atmospheric self-gravity, triggering runaway gas accretion and inflating the
planet into a gas giant. Concurrent with these processes are the inward 
migration of the planet and the dispersal of the gas disc (driven by
viscous accretion in our model).
} 
 \label{fig:overview}
\end{figure*} 

\section{Introduction}
Gas giant planets can only form when there is still enough gas in their
natal environments, i.e., when circumstellar discs are young
and gas-rich. The core accretion paradigm of gas giant formation
plays out in three phases \citep{pollack96}. Phase 1: a rocky core 
coagulates from disc solids. Phase 2: a gaseous envelope forms around
the core, fed from the ambient disc at a rate regulated by internal cooling 
and contraction of the envelope.
Phase 3: the planet inflates into a gas giant
as the gas accretion rate onto the core
``runs away'' in response to the envelope's self-gravity.

Of these three phases, perhaps the first is the least understood. 
There is a need for gravitational focussing of collisions between solid
particles, as timescales
for coagulating a Jupiter-breeding core 
(having $\gtrsim$ a few Earth masses) at the stellocentric
distances where Jupiters are found
(a few AU) are orders of magnitude longer than the
1--10 Myr lifetimes of gas discs---if collision cross-sections
are only geometric and not enhanced by gravity \citep{Goldreich04}. 
How much collisions are focussed depends on much particle velocity
dispersions are damped, which in turn depends on particle
size distributions. These factors are not known.
The problem is tied up
with the longstanding mystery of planetesimal formation
\citep{Chiang10}.

``Pebble accretion'' makes inroads on the problem of core assembly
by exploiting the ability of seed cores
to attract particles small enough to have their
velocity dispersions damped by aerodynamic drag from the ambient gas disc (for a comprehensive review,
see \citealt{Ormel2017}). 
Without addressing the question of the origin of the seed
core (seeds as low in mass as $\sim$10$^{-3} M_\oplus$ have been assumed),
the theory of pebble accretion points out that for particles whose
aerodynamic stopping times are comparable to orbital
times---``pebbles'' with order-unity Stokes numbers---the
accretion cross-section can approach its maximum value set by 
the Hill sphere of the seed core. Assembly times
of Jupiter-breeding cores have been reported to range from
$\sim$10$^4$ yr \citep{Lambrechts12} to $\sim$10$^6$ yr 
\citep{Lambrechts14}. Pebble accretion has been hailed
as a key ingredient in understanding various architectural features
of planetary systems, both in Solar and extrasolar contexts, 
including: the dichotomy between inner terrestrial planets
and outer giants \citep{Morbidelli15, Levison15};
the preponderance of gas giants
at a few AU and of
ice giants beyond \citep{Bitsch15a,Bitsch15b}; 
the orbital period distribution of warm Jupiters \citep{Ali-Dib17};
and the mass distribution of planets in the TRAPPIST-1 system \citep{Ormel_trappist17}.

Although pebble accretion is touted for its speed,
it is perhaps less appreciated that it can be lossy.
The same aerodynamic drag that brings pebbles into contact
with cores also sweeps them past, as part of the background
radial drift of solids from the outer to the inner disc \citep[e.g.,][]{weiden77}. Particles that are most easily
accreted, with Stokes numbers near unity, also drift the fastest
toward the star. Factoring in this background flow---tallying
how many Earth masses flow under the bridge for every Earth mass
netted---is one aim of this paper. 
We study how pebble
accretion plays out in a global (1D in radius), fully
time-dependent model, one that 
balances the disc's finite mass budget (i.e., keeps
track of the various sources and sinks of solid mass)
while growing a core.
We consider pebble sizes ranging from 0.01--1 cm, like those
probed by mm--cm wavelength images of protoplanetary discs
(e.g., \citealt{ALMA15}; \citealt{Andrews16}), and
near the maximum size that might be achievable
by particle-particle
sticking \citep[][their section 4]{Chiang10}. 
Aside from specifying a few model inputs such
as the disc's initial solid mass, its gas mass,
and its Shakura-Sunyaev $\alpha$,
we make no assumption about the regime of pebble
accretion that a protocore finds itself in,
letting system parameters
dictate which of the myriad cases outlined by
\citet{Ormel10} is appropriate at any given time---in this regard,
our approach is more physically conservative than other treatments.
In addition to solving the equations of pebble accretion,
the model also accounts for gas accretion onto cores, 
and orbital migration of planets by disc torques. 
Our goal is to assess more
realistically the prospects of pebble accretion at forming
various kinds of planets---Earths, 
Neptunes, 
and, in particular, Jupiters---in discs 
resembling those observed. 
We focus our attention on the growth of a single core 
to assess whether a given disk can create even a
single giant planet.
We describe our model in section \ref{sec:method}, present results in section \ref{sec:results}, and summarize and look to the future in section \ref{sec:sum}.

\section{Model}
\label{sec:method}

We construct a global (1D in disc radius) model of planet formation. 
Figure \ref{fig:overview} 
presents a pictorial overview. 
The model tracks the evolution of a viscous gas disc (section \ref{gdisk}); the radial transport of solid particles of fixed size (section \ref{sdisk}); the assembly of a single rocky core by pebble accretion (section \ref{core}); the lowering
of disc gas density near the planet's orbit due to gap
opening (section \ref{gap});
how the core stops accreting solids (section \ref{gap_dust}); how
cores that have stopped accreting solids subsequently accrete
nebular gas (sections \ref{atm} and \ref{run}); and migration of the nascent planet (section \ref{mig}). Readers interested in a bare-bones technical summary of the underlying equations and
values of input parameters can jump to section \ref{tech}.  Table \ref{tab:var} lists our symbols and their meanings.

\begin{table*} 
\renewcommand\thetable{1} 
\caption{Model Parameters}
\label{tab:var}
\begin{tabular}{llp{6cm}ll}
\hline
\hline
Used for & Symbol & Description & Values & Reference\\
\hline
Gas disc & $\alpha$ & Viscosity parameter / turbulent Mach number & $10^{-4}$, $10^{-3}$, $10^{-2}$ & Section \ref{gdisk}\\
&$a_1$ & Characteristic disc radius & 10, 100 AU & Section \ref{gdisk}\\
&$M\sub{disc,g}(0)$ & Initial mass of gas disc & $10$, 100 $M\sub{J}$ & Section \ref{gdisk}\\
&$T$ & Disc temperature & $260\,{\rm K}\times (a/1 \,{\rm AU})^{-1/2}$ & Section \ref{gdisk}\\
Solid disc&$a\sub{in}$ & Inner radius of the solid disc & 0.01 AU & Section \ref{sdisk}\\
&$a\sub{out}(0)$ & Initial outer radius of the solid disc & $a_1$ & Section \ref{sdisk}\\
&$s$ & Pebble radius & 0.01, 0.1, 1 cm & Section \ref{sdisk}\\
&$Z$ & Solid-to-gas mass ratio $M\sub{disc,s}(0)/M\sub{disc,g}(0)$ & 0.003, 0.009, 0.031 & Section \ref{sdisk}\\
Core & $a(0)$ & Initial core orbital radius & 0.1, 0.3, 1, 3, 10, 30 AU & Section \ref{tech}\\
&$M_{\rm core}(0)$ & Initial seed core mass & 0.01 $M_\oplus$ & Section \ref{tech}\\
&$M_{\rm gas}(0)$ & Initial planet gas mass & 0 & Section \ref{tech}\\
\hline
\multicolumn{5}{p{0.9\textwidth}}{Notes: (a) In selecting model parameters, we ensure that the initial core orbital radius $a(0)$ is strictly less than the characteristic disc radius $a_1$. In total, we have 540 unique parameter combinations. (b) Initial solid disc masses $M_{\rm disc,s}(0) = Z M_{\rm disc,g}(0) \in \{ 10, 30, 100, 300, 1000 \} M_\oplus$ (only 5 unique values of $M_{\rm disc,s}(0)$ corresponding to 6 unique combinations of $M_{\rm disc,g}(0)$ and $Z$).}
\end{tabular}
\end{table*}

\subsection{The Gas Disc} \label{gdisk}

To model the gas surface density $\Sigma_{\rm g}$ as a
function of radius $a$ and time $t$,
we adopt the similarity solution for an isolated, viscously spreading disc
\citep{lynden_bell74,hartmann98}:
\begin{equation} \label{sig_g}
    \Sigma\sub{g}(a,t) = \frac{Ca_1}{3\pi \nu_1 a} \mathcal{T}^{-3/2} \text{exp} \left[ \frac{-a/a_1}{\mathcal{T}} \right] 
\end{equation} 
where $a_1$ is a fiducial radius,
$\nu_1 = \nu(a_1)$ is the kinematic viscosity at $a_1$, 
$\mathcal{T} = t / t\sub{s}+1$, and 
\begin{equation} \label{ts}
    t\sub{s} = \dfrac{a_1^2}{3\nu_1}
\end{equation}
is a measure of the viscous diffusion time at $a_1$. Equation (\ref{sig_g})
presumes that the disc viscosity scales as
\begin{equation}\label{eq:nu}
  \nu = \alpha c\sub{s} H \propto a^1
\end{equation}
for sound speed $c\sub{s}$, scale height $H = c\sub{s}/\Omega$,
orbital angular frequency $\Omega$, 
and dimensionless viscosity
parameter $\alpha$ \citep{shakura73}.
The scaling of (\ref{eq:nu}) follows, in turn, from
an assumed temperature profile
\begin{equation} \label{eq:T}
    T = T_0 \Big(\frac{a}{a_0}\Big)^{-1/2}
\end{equation}
for $a_0 = 1$ AU and $T_0 = 260$ K.
The normalization constant $C$ depends on an assumed initial gas mass
of the disc $M_{\rm disc,g}(0)$. Given (\ref{eq:T}), the disc aspect
ratio is
\begin{equation} \label{eq:H}
H/a = 0.032 \left( \frac{a}{a_0} \right)^{1/4} \,.
\end{equation}
In sum, there are three free
parameters: $M_{\rm disc,g}(0)$,
$\alpha$, and $a_1$ (equivalently, $t\sub{s}$). 
The ranges of these and
other variables are given in Table \ref{tab:var}.

\subsection{The Solid Disc} \label{sdisk}
We follow the contraction of the solid disc as its
constituent particles drift inward by gas drag.
In inertial space, particles travel radially
inward at a speed
\begin{equation} \label{vdrift}
v\sub{drift}= \frac{3\nu}{2a} \frac{1}{1+\tau^2} + 2 v\sub{hw} \frac{\tau}{1+\tau^2}
\end{equation}
where $3\nu/(2a)$ is the steady-state viscous gas velocity,
$v\sub{hw}$ is the azimuthal headwind velocity experienced
by particles in pressure-supported gas,
and $\tau = \Omega t_{\rm stop}$
is the dimensionless particle
stopping time (a.k.a.~Stokes number). 
Equation (\ref{vdrift}) generalizes equation (13) of
\citet{Chiang10}.
The headwind velocity is
\begin{equation} \label{vhw}
v\sub{hw} = -\frac{c\sub{s}^2}{2\Omega a}\pd{\log P\sub{g}}{\log a},
\end{equation}
where the gas pressure $P\sub{g}$
and density $\rho\sub{g}$ are approximated as
\begin{equation}
    P\sub{g} = \rho\sub{g} c\sub{s}^2 = \dfrac{\Sigma\sub{g}}{H}c\sub{s}^2 \,.
\end{equation}
The particle stopping time is
\begin{equation} \label{tau}
\tau=
\begin{cases} 
      \dfrac{\Omega \rho\sub{s} s}{\rho\sub{g} c\sub{s}} & s<\frac{9}{4}\lambda  \hspace{.75cm} (\text{Epstein drag})\\
      \dfrac{4 \Omega \rho\sub{s} s^2 }{9 \rho\sub{g} c\sub{s} \lambda} & s>\frac{9}{4} \lambda  \hspace{.75cm} (\text{Stokes drag})
\end{cases}
\end{equation}
\noindent where $s$ is the particle radius,
$\rho\sub{s} = {\rm1\,g/cm}^3$ is the internal density of a single particle, and $\lambda = 4 \times 10^{-9}/\rho\sub{g}$
is the gas mean free path in cgs units.
For simplicity we take $s$ to be strictly constant in a given model
(cf.~\citealt{Ormel12} who relax this assumption).

The radial transport of solids---i.e., the
evolution of solid surface density
$\Sigma_{\rm s}(a)$ with $t$---is
solved using a simple Lagrangian scheme.
At $t=0$,
the solid disc, of total mass $M_{\rm disc,s}(0)$,
is divided into 1000 concentric rings that are 
logarithmically spaced from $a_{\rm in} = 0.01$ AU to
$a_{\rm out}(0) = a_1$. 
Mass is assigned to each ring such that the initial solid surface
density scales as $a^{-1}$ (following the viscously
relaxed portion of the gas disc; section \ref{gdisk}).
Each ring conserves its mass
(the mass lost to pebble accretion onto the planetary core
is negligible) 
but has its radial boundaries evolved according to
equation (\ref{vdrift}). The evolution of the ring boundaries
is solved using a
fourth-order Runge-Kutta scheme from $t = 10^{-3}$ Myr to
100 Myr over 5000 logarithmically spaced timesteps. 
The solid surface density of each ring is simply the 
ring mass divided by the (evolving) ring area. 
A ring whose inner boundary crosses inside $a_{\rm in}$ has
its inner boundary held at $a_{\rm in}$ and its surface density
fixed thereafter; once the ring's outer boundary crosses
$a_{\rm in}$, the ring is removed from the calculation.
To reduce numerical noise when calculating $\Sigma_{\rm s}$ for a
given orbital radius $a$ of the core, we use cubic splines to
interpolate over the solid surface
densities of up to 10 rings inside $a$, plus up to 10 rings outside (hitting 
disc boundaries can limit the number of rings used in the interpolation). 
Because the innermost ring is handled differently 
as per the above, it is not used in any interpolation, and if the core
falls within the innermost ring, its local $\Sigma_{\rm s}$ is simply
that of that ring.

\begin{figure}
    \centering
    \includegraphics[width = \columnwidth]{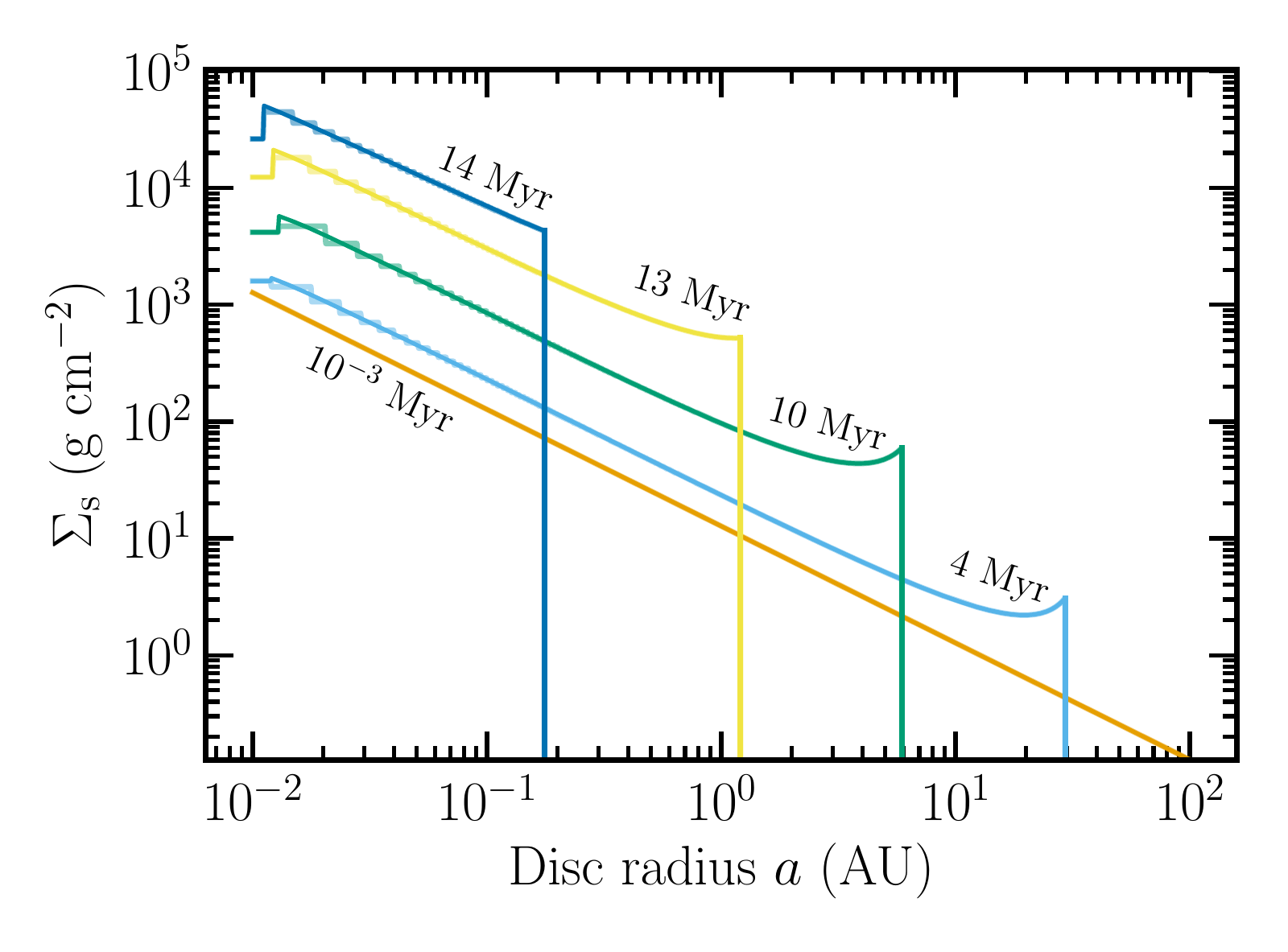}
    \caption{Surface density of solids, evolved according to our Lagrangian
    ring scheme (section \ref{sdisk}), for a disc with $\alpha = 10^{-4}$,
    $a_1 = 100$ AU, $s = 0.01$ cm, $M_{\rm disc,g}(0)= 100 M_{\rm J}$, and
    $M_{\rm disc,s}(0) = 300 M_\oplus$ 
    (note that these
    parameters are not used in subsequent case examples
    but are chosen to most cleanly illustrate radial drift).
    Solid surface densities are interpolated over raw simulation results
    (step-like curves).
    For these model parameters, the surface density at a fixed location---say
    $a = 1$ AU---rises with time in a ``particle pile-up''
    before the entire solid disc sweeps by.
}
    \label{fig:sdisk0}
\end{figure}

A sample evolution of $\Sigma_{\rm s}(a,t)$
is shown in Figure \ref{fig:sdisk0}.
The solid disc introduces two free parameters, the 
particle size $s$ and the initial total solid mass
$M_{\rm disc,s}(0) \equiv Z M_{\rm disc,g}(0)$
(see Table \ref{tab:var}).

\begin{figure*}
    \centering
    \includegraphics[width=\textwidth]{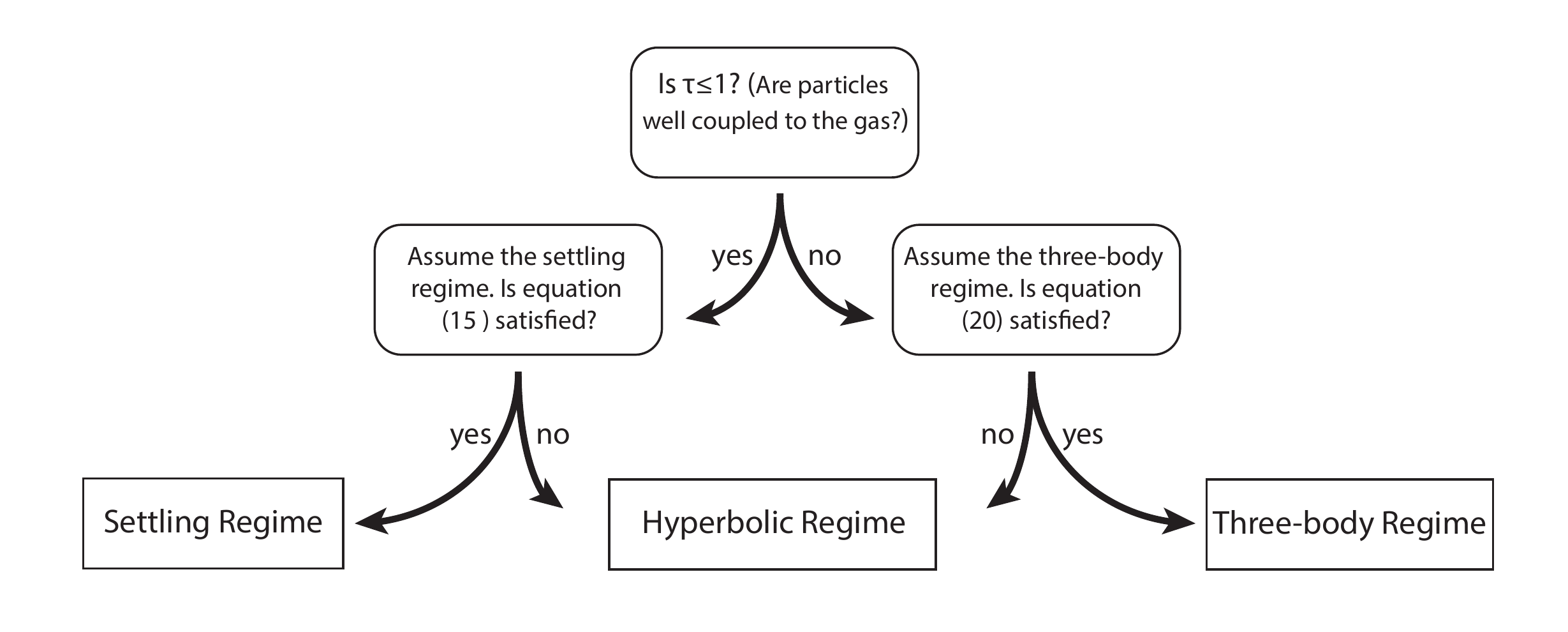}
    \caption{Flowchart for choosing the appropriate pebble accretion regime. In the case where the hyperbolic regime is returned, we perform additional {\it a posteriori} checks given by equations (\ref{eq:hyp_check_1}) and (\ref{eq:hyp_check_2}). In practice, for our parameters, pebble accretion is nearly always in the settling regime.}
    \label{fig:flow}
\end{figure*}

\subsection{Core Formation by Pebble Accretion} \label{core}

To grow cores starting from an assumed seed mass of
$M_{\rm core}(0) = 10^{-2} M_\oplus$, we follow the pebble accretion
prescriptions of \citet[][OK10]{Ormel10}, augmenting their formulae
to account for gas turbulence (see also \citealt{Ormel18} and
\citealt{Rosenthal18} who
provide a more general treatment of stochasticity in pebble accretion). 
A core embedded in a disc of solids accretes mass at a rate
\begin{equation}
    \label{core_eq}
    \dot{M}\sub{core} = 2\Sigma_{\rm s}R_{\rm acc}v_{\rm acc} \times
    \min (1, R_{\rm acc}/H\sub{s})
\end{equation} 
where particles that come within a distance $R_{\rm acc}$ of the core,
moving at velocity $v\sub{acc}$ relative to the core, are accreted.
According to (\ref{core_eq}), 
when $R_{\rm acc}$ (the accretion ``cross section'' or impact parameter)
is larger than the vertical thickness of the disc of solids,
\begin{equation} \label{hs}
    H\sub{s} = H \sqrt{\dfrac{\alpha}{\alpha+\tau}}
\end{equation}
\citep{Youdin07}, then the particles effectively comprise a 2D sheet; otherwise, the thickening of the particle layer due to turbulence
reduces the pebble accretion rate by a factor $R\sub{acc}/H\sub{s}$.
For our parameter space,
$\min H_{\rm s} = \min H\sqrt{\alpha/\tau} \simeq H/40$
(characterizing only for a few models);
this thickness is comparable to the
physical minimum imposed by Kelvin-Helmholtz shear turbulence
($\sim$$H^2/a \simeq H/30$; e.g.,
\citealt{Lee10a,Lee10b,Rosenthal18}).

OK10 identify three regimes---settling, three-body, and
hyperbolic---each having their own forms for
$R_{\rm acc}$ and $v\sub{acc}$.
These formulae, modified for gas turbulence,
are provided in the subsections below.

We erect local Cartesian axes centred on the core
at orbital radius $a$,
where $x$ increases radially outward 
and $y$ advances in the direction
of the core's (assumed circular) orbital motion.
Relative to the core, the particles have mean velocity
(cf.~equation \ref{vdrift})
\begin{subequations} \label{eq:vxvy}
    \begin{align}
        v_x &= -\frac{2v_{\rm hw}\tau}{1+\tau^2} -\frac{3\nu}{2 a}\frac{1}{1+\tau^2} - \dot{a}, \label{eq:vx}\\
        v_y &= - \frac{v_{\rm hw}}{1+\tau^2} + \frac{3\nu}{2a} \frac{\tau}{2(1+\tau^2)} - \frac{3}{2}\Omega x \label{eq:vy}
    \end{align}
\end{subequations}
where the terms involving $\tau$ are due to drag in sub-Keplerian, viscously accreting 
gas (generalizations of equations 13 and 14 of \citealt{Chiang10}), $\dot{a}$ is the core's radial migration velocity (section \ref{mig}), 
and the term proportional to $x$ is from Keplerian shear.
In addition to this mean relative velocity, gas turbulence
imparts randomly oriented velocities to the particles of magnitude
\begin{equation}
    v_{\rm turb} = c_{\rm s} \sqrt{\frac{\alpha}{1+\tau}}
    \label{eq:vturb}
\end{equation}
\citep[see][their equation 13]{Youdin07}.
The total relative speed between the core and the particles
is calculated by adding in quadrature the mean 
$\sqrt{v_x^2+v_y^2}$ and the fluctuation $v_{\rm turb}$.

\subsubsection{Settling Regime}
\label{sssec:settling}

In the settling regime, particles are well-coupled to the gas
($\tau \leq 1$).
A particle is assumed to accrete onto the core if, upon 
approaching the core with velocity $v_{\rm acc}$ and
acquiring a specific impulse (``kick'') $\Delta v$ after the encounter,
its trajectory is deflected by an order-unity
angle. OK10 write this condition as $\Delta v \sim v_{\rm acc}/4$.

For simplicity we use an expression for
$v_{\rm acc}$ valid
in the limit $\tau \ll 1$,
dropping mean velocities that are typically smaller
than the azimuthal headwind velocity
(cf.~equations \ref{eq:vxvy} and \ref{eq:vturb}):
\begin{equation}
   v_{\rm acc} \sim \sqrt{\left(v_{\rm hw} + \frac{3}{2}\Omega R_{\rm acc}\right)^2 + \alpha c_{\rm s}^2} \,.
   \label{eq:vacc_settle}
\end{equation}
The settling regime is further defined by the condition
that the particle stopping time
be shorter than the encounter time:
\begin{equation} \label{eq:settle_condition}
\tau/\Omega < R_{\rm acc}/v_{\rm acc} \,.
\end{equation}
Under this assumption, the kick velocity is
\begin{equation} \label{eq:kick}
\Delta v \sim \frac{GM_{\rm core}}{R_{\rm acc}^2}\frac{\tau}{\Omega} \,,
\end{equation}
i.e., the particle is assumed to attain
terminal velocity during the encounter
(with the gravitational force from the core balancing gas drag).
Setting $\Delta v$ equal to $v_{\rm acc}/4$, and re-writing
in Hill's units, we have
\begin{equation}
 \frac{9}{4}b^6 + 3\zeta b^5 + (\zeta^2 + \alpha f_P^{-1}\mu^{-1/3}\zeta)b^4 - 144\tau^2 = 0
    \label{eq:b_settle}
\end{equation}
which we solve for $b = R_{\rm acc}/R_{\rm Hill}$.
Here $R_{\rm Hill} = \mu^{1/3} a$, $\mu = M_{\rm core}/(3M_\star)$,
$M_\star$ is the stellar mass,
$\zeta = v_{\rm hw}/v_{\rm Hill}$, 
$v_{\rm Hill} = \Omega R_{\rm Hill}$,
and $f_P = -(1/2)\partial \log P_{\rm g}/\partial \log a$.
Having solved (\ref{eq:b_settle}) for $R_{\rm acc}$,
we insert into (\ref{eq:vacc_settle}) to evaluate
$v_{\rm acc}$. Our equation (\ref{eq:b_settle})
is analogous to equation (27) of OK10 except that
our accounting for turbulence in $v_{\rm acc}$
has resulted in a higher-order polynomial.

\subsubsection{Three-Body Regime}
\label{sssec:3body}

In the three-body regime, particles execute trajectories
in and around the Hill sphere that are perturbed only slightly
by gas drag ($\tau \gg 1$). 
Accretion in this regime
should be similar to that of a 
sub-Hill disc
and its chaotic Hill sphere dynamics (\citealt{Goldreich04},
their sections 3.3 and 3.4; our 2D/3D correction factor
$\min (1,R_{\rm acc}/H_{\rm s})$ in equation \ref{core_eq}
distinguishes between
their ``not very thin'' and ``very thin'' discs).
Following OK10, we take
\begin{equation}
    v_{\rm acc} = 3.2 v_{\rm Hill}
    \label{eq:vacc_3b}
\end{equation}
and
\begin{equation}
    b = 1.7\alpha_{\rm core}^{1/2} + \frac{1}{\tau}
    \label{eq:b_3b}
\end{equation}
where $\alpha_{\rm core} \equiv R_{\rm core}/R_{\rm Hill}$ (not to 
be confused with the gas turbulent Mach
number $\alpha$), and
the term $1/\tau$ is OK10's empirical correction for how gas drag 
enhances the accretion cross section (and which
can be justified using an energy argument; \citealt{Rosenthal18}).

In the quasi-Hill sphere dynamics assumed by the three-body regime,
the dominant velocity with which a particle approaches the core
is set by the Kepler
shear term $(3/2)\Omega x \sim \Omega R_{\rm Hill} = v_{\rm Hill}$
in equations (\ref{eq:vxvy})--(\ref{eq:vturb}).
We write this defining condition as follows,
assuming for simplicity that $\tau \gg 1$, and
dropping the terms proportional to the 
viscous gas velocity $3\nu/(2a)$
and the migration velocity $\dot{a}$ which are
typically negligible compared to $v_{\rm Hill}$:
\begin{equation}
\sqrt{4v_{\rm hw}^2/\tau^2 + \alpha c_{\rm s}^2/\tau} < v_{\rm Hill} 
\end{equation}
which re-written in Hill's units reads
\begin{equation}
    \tau^2 - \alpha f_P^{-1}\mu^{-1/3}\zeta\tau-\zeta^2 > 0.
    \label{eq:3b_hyp_bdy}
\end{equation}
When $\alpha=0$, this reduces to the same condition $\tau \gtrsim \zeta$
used by OK10 to demarcate the three-body regime (see their Figure 7).

\subsubsection{Hyperbolic Regime}
\label{sssec:hyperbolic}

In the hyperbolic regime, particle-core encounters are fast---faster
than the particle stopping time:
\begin{equation} \label{eq:hyp_check_1}
R_{\rm acc}/v_{\rm acc} < \tau/\Omega
\end{equation}
in violation of the settling condition (\ref{eq:settle_condition}),
and also faster than the time to cross the Hill sphere
at the Hill velocity:
\begin{equation} \label{eq:hyp_check_2}
R_{\rm acc}/v_{\rm acc} < R_{\rm Hill}/v_{\rm Hill} = 1/\Omega
\end{equation}
in violation of the Hill sphere dynamics assumed by the three-body
regime. The encounter therefore unfolds in a classic
two-body fashion with a gravitationally focused accretion
cross section of
\begin{align}
    b &= \alpha\sub{core} \sqrt{1+\left(\frac{v_{\rm esc}}{v\sub{acc}}\right)^2} \nonumber \\
    &= \alpha_{\rm core} \sqrt{1+\frac{6}{\alpha_{\rm core}(v_{\rm acc}/v_{\rm Hill})^2}}
    \label{eq:b_hyp}
\end{align}
where $v_{\rm esc}$ is the escape velocity from the core surface
and $v_{\rm esc}/v_{\rm Hill} = \sqrt{6/\alpha_{\rm core}}$.
In general, the relative velocity $v_{\rm acc}$ equals 
$\sqrt{v_x^2 + v_y^2 + v_{\rm turb}^2}$; but in evaluating
(\ref{eq:vy}) for $v_y$, we may drop the Kepler shear term
since $b \ll 1$ and $\zeta = v_{\rm hw}/v_{\rm Hill} > 1$
in the hyperbolic regime. Thus
\begin{equation}
\begin{split}
    \frac{v_{\rm acc}}{v_{\rm Hill}} \sim \left[ \left(\dfrac{2 \zeta \tau}{1+\tau^2} + 
    \frac{3}{2}\frac{\alpha \zeta f_P^{-1}}{1+\tau^2}+\dfrac{\dot{a}}{v_{\rm Hill}}\right)^2 \right. \\ 
    + \left. \left( \dfrac{\zeta}{(1+\tau^2)} - \dfrac{3}{2} \frac{\alpha\zeta f_P^{-1} \tau}{2(1+\tau^2)} \right)^2 + \frac{\alpha \zeta \mu^{-1/3}}{f_P (1+\tau)}\right]^{1/2} \,.
\end{split}
\end{equation}
This matches equation (29) of OK10 when $\alpha = 0$
and $\dot{a} = 0$.

We summarize in Figure \ref{fig:flow} our decision tree for
choosing the appropriate regime of pebble accretion as we track
the growth of a core.
When the hyperbolic regime is selected, we 
verify {\it a posteriori} that the inequalities
(\ref{eq:hyp_check_1}) and (\ref{eq:hyp_check_2}) are 
satisfied.
In practice, we find for our model parameters
that pebble accretion is almost always in the settling regime
(cf.~Figure \ref{fig:Mcore_v_t_compLJ14} which uses a disjoint 
set of model inputs and which shows some cases of hyperbolic
accretion).
As noted in section \ref{sdisk},
in evaluating $\Sigma\sub{s}$ in the accretion rate (\ref{core_eq}),
we interpolate using cubic splines in the vicinity of the core.

\subsection{Opening a Gap in the Gas Disc}
\label{gap}

A core exerts Lindblad torques on the gas disc that open
an annular gap about its orbit. The gas surface density
at the position of the planet,
$\Sigma_{\rm g,planet}$, is related to the local unperturbed value 
$\Sigma_{\rm g}$ by
\begin{equation} \label{sigma_gp}
    \frac{\Sigma_{\rm g}}{\Sigma_{\rm g,planet}} - 1 = 0.043 \left(\frac{M_{\rm planet}}{M_\star}\right)^2 \left(\frac{H}{a}\right)^{-5}\alpha^{-1} 
\end{equation}
(\citealt{dong17}; see also \citealt{kanagawa15} and \citealt{fung14}).
Here $M\sub{planet} = M_{\rm core} + M_{\rm gas}$ is the total planet mass, including not only the rocky core
but also any gas envelope it carries (accretion of
gas onto the core is treated in section \ref{atm}),
while $M_\star = M_\odot$ is the host star mass.

Depressing the gas surface density at the position of the planet
has two effects. First, it reduces---very slightly---the rate
at which the solid core accretes a gas envelope (see the weak
dependence on $\Sigma_{\rm g,planet}$ in equation \ref{dustfree}).
Second and more significantly, once the gas gap becomes deep
enough, orbital migration of the planet is expected
to slow from Type I to Type II (section \ref{mig}).

\subsection{Halting Pebble Accretion} \label{gap_dust}
As a planet grows in mass and opens a deeper gap in the gas disc,
local pressure gradients near gap edges may trap
particles there. \citet[][LJ14]{Lambrechts14}
term the core mass above which solids
can no longer drift onto the core the ``pebble isolation mass.''
In hydrodynamics simulations, the pebble isolation
mass is found to scale as the thermal mass (that for which
surface density perturbations excited by the planet
become non-linear):
\begin{align} \label{eq:peb_iso}
M_{\rm peb,iso} &= 5 M_\oplus \left( \frac{H/a}{0.03} \right)^{3} \left[  0.34 \left( \frac{-3}{\log_{10} \alpha} \right)^4 +0.66 \right]
\end{align}
(\citealt{bitsch18}; see also equations 33 and 36 of \citealt{dipierro17}; \citealt{Ataiee18};
\citealt{rosotti16}; LJ14).

In our model,
a core stops accreting particles either when 
(a) $M_{\rm planet} > M_{\rm peb,iso}$, or 
(b) the entire disc of solids has drifted past the core
(section \ref{sdisk}). Outcome (b) will prove common.

\subsection{Gas Accretion onto Cores} \label{atm}

\begin{figure}
    \centering
    \includegraphics[width=\columnwidth]{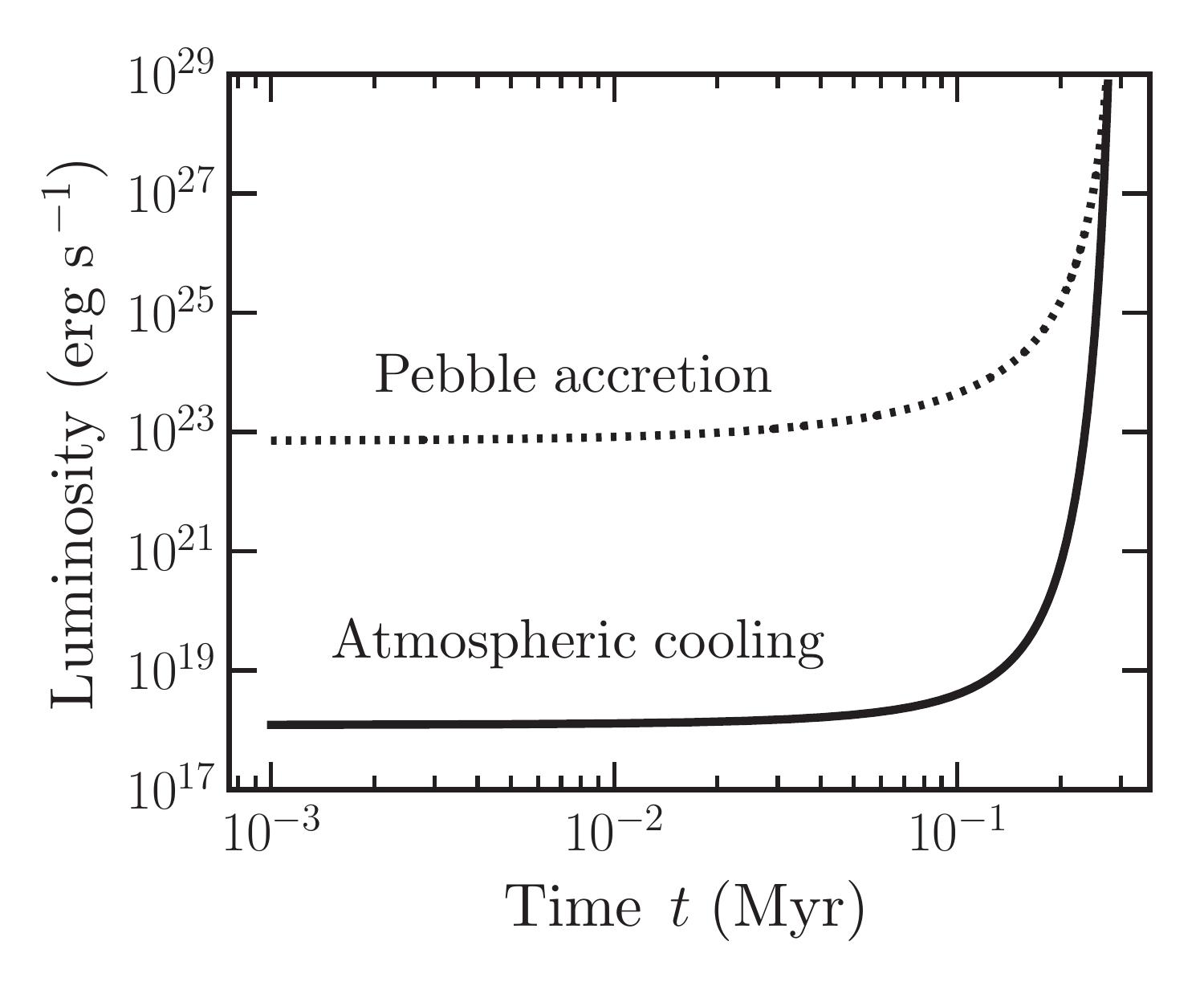}
    \caption{Pebble accretion luminosities tend to overwhelm
      atmospheric cooling luminosities, preventing gas from accreting onto
      solid cores until after pebbles have stopped accreting. The
      pebble accretion luminosity (dotted curve) is computed as $GM_{\rm core}
      \dot{M}_{\rm core} / R_{\rm core}$, where $\dot{M}_{\rm core}$
      is computed using (\ref{core_eq}) and $R_{\rm core}$ derives
      from $M_{\rm core}$ assuming a fixed bulk density of
      $5\,{\rm g\,cm^{-3}}$.
      The core mass is initially set to $0.01 M_\oplus$ at a time of
      $10^{-3}$ Myr, and accretes pebbles with
      size $s = 0.1$ cm at a nearly constant $a = 1$ AU, 
      in a disc for which $\alpha = 10^{-4}$, $a_1 = 100$ AU,
      and the initial gas and solid masses are 
      $M_{\rm disc,g}(0) = 10 M_{\rm J}$ and $M_{\rm disc,s}(0) = 100 M_\oplus$
      (these same disc parameters are adopted for Figure  \ref{fig:wj}).
The core grows until the solid disc drifts entirely inside of the core's orbit,
at which point $M_{\rm core} \simeq 3 M_\oplus$.
To compute the atmospheric cooling
luminosity (solid curve), we use \citet[][their equations 8 and afterward]{Lee15},
with $T_{\rm rcb}$ set equal to $260$ K,
opacity constants appropriate to their section 2.2.2 for dust-free and
gas-rich discs, and the gas-to-core mass ratio fixed at GCR = 0.01; assuming a higher GCR only lowers the cooling luminosity and strengthens the point of this plot.}
    \label{fig:lum1}
\end{figure}

Cores accrete as much gas as can cool---which it cannot 
while pebble accretion is ongoing, as the rate of accretional
heating tends to exceed the rate of envelope cooling by orders
of magnitude (Figure \ref{fig:lum1}). 
We therefore start gas accretion onto solid cores
only after the cores have
finished accreting pebbles (section \ref{gap_dust}).

We adopt the scaling relation
for atmospheric growth derived by \citet[][their equation 24, derived for temperatures between 100 K and 800 K]{Lee15},
accounting for the weak dependence on nebular surface density
\citep[][their Figure 4; we have verified that this density dependence
applies at different distances]{Lee16}:
\begin{align} 
\label{dustfree}
     \frac{M_{\rm gas}}{M_{\rm core}} &= 0.20 \, \Big(\dfrac{\Delta t}{0.1 \, \text{Myr}}\Big)^{0.4} \Big( \dfrac{500 \, \text{K}}{T} \Big)^{1.5} \nonumber \\
       &\times \Big( \dfrac{M\sub{core}}{5\,M_\oplus} \Big)^{1} \Big( \dfrac{\Sigma\sub{g,planet}}{0.03\,\Sigma \sub{mmen}} \Big)^{0.12} 
\end{align}
where $M_{\rm gas}$ is the planetary gas mass,
$\Delta t$ is the elapsed time since the onset
of gas accretion, $T$ is the nebular temperature,
and $\Sigma\sub{mmen}$ is the surface
density of the minimum-mass extrasolar nebula
\citep{mmen}:
\begin{equation}
    \Sigma \sub{mmen} = 4\times10^{5} \left( \dfrac{a}{0.1 \,{\rm AU}}\right)^{-1.6} \:{\rm g/cm}^2 \,.
\end{equation}
Equation (\ref{dustfree}) drops the contribution from
any dust to the gas opacity; this is a plausible assumption
given grain coagulation, settling, and ablation
\citep{ormel14,brouwers18}.\footnote{\citet{Lambrechts17}
argue otherwise, that dust from disc gas
is continuously brought to the planet's outer envelope.
We therefore also experiment by replacing equation (\ref{dustfree})
with an analogous formula based on an opacity that
includes dust \citep[][their equation 20]{Lee15}.
None of the conclusions of this paper
changes quantitatively; more details on this separate set
of experiments are given in the Results section,
footnote \ref{foot:dusty}.\label{foot:predusty}}
We also ignore for simplicity the effects of super-solar 
gas-phase metallicity, which can hasten the rate of cooling and gas accretion, but only for metallicities $Z \gtrsim 0.5$ (e.g., \citealt{Venturini15}, their Figure 1; see also \citealt{Venturini17}).

Equation (\ref{dustfree}) further presumes that the bulk of the
envelope mass is centrally concentrated near the core, i.e.,
the adiabatic index $\gamma_{\rm ad} < 4/3$ in the inner convective
zone. This assumption is valid when envelope temperatures $> 2500$ K so
that H$_2$ dissociates.  Only cores more massive than about
$0.5 M_\oplus$ can gravitationally retain such gas, and so we apply
equation (\ref{dustfree}) 
to grow atmospheres only for cores which
exceed $0.5 M_\oplus$ at the end of the pebble accretion
phase.\footnote{We have also determined, following the methodology of \citet{Lee14}, that core masses $< 0.5 M_\oplus$ take longer than
  $\sim$10 Myrs to accrete even 1\%-by-mass envelopes.}
For such cores, we combine the time derivatives of equation
(\ref{dustfree}) with the planet's time-varying orbital distance
(section \ref{mig}) and the disc's time-varying surface density
(section \ref{gdisk}) to track increments in the planet's gas mass as
ambient nebular conditions change.  In taking the time derivative of
(\ref{dustfree}), we ignore terms depending on $dT/dt$ and
$d\Sigma_{\rm g,planet}/dt$ for simplicity. We also check that
the envelope mass does not exceed the value that it would have
if it were isothermal at the disc temperature; this is the maximum
mass to which the atmosphere can grow, as gas cannot cool past this
point \citep[see][their Figure 4]{Lee15}. If the envelope mass exceeds 
this isothermal bound, we halt gas accretion.

\subsection{Runaway Gas Accretion and the Formation of Jupiters} \label{run}

Once the gas envelope mass becomes comparable to the underlying
core mass, the envelope's self-gravity becomes significant.
Self-gravitating envelopes demand larger luminosities
to sustain hydrostatic equilibrium; as gas continues to pile on,
the cooling timescale shortens
catastrophically, and the envelope grows in a runaway fashion 
\citep[e.g.,][]{pollack96}.
We crudely model runaway using a step function:
once $M_{\rm gas}/M_{\rm core} \geq 0.5$, we boost 
$M_{\rm planet} = M_{\rm gas} + M_{\rm core}$ to $1 M_{\rm J}$, 
or we add to $M_{\rm gas}$ the total mass in the gas disc outside the planet's orbit,
whichever option yields a smaller $M_{\rm planet}$.
The latter option is not to be taken literally
(i.e., we do not literally set $\Sigma_{\rm g}$ to zero outside
the planet's orbit), but is merely a rough proxy
for mass conservation.

\subsection{Disc-Driven Migration} \label{mig}
We define a ``deep'' gap as one for which
$\Sigma_{\rm g,planet}=1/10$ the value of the unperturbed
surface density $\Sigma_{\rm g}$.
From (\ref{sigma_gp}), deep gaps are opened by
planet masses exceeding
\begin{align} \label{eq:gap_mass}
M_{\rm Type\,II} &= 14.5 M_\star \alpha^{1/2} \left(\frac{H}{a}\right)^{5/2} \nonumber \\
& = 8 M_\oplus \left( \frac{\alpha}{10^{-4}} \right)^{1/2} \left( \frac{H/a}{0.03} \right)^{5/2} \,.
\end{align}
Planets for which $M_{\rm planet} < M_{\rm Type\,II}$ are transported
radially according to the Type I migration rate:
\begin{equation} \label{eq:type1}
    \dot a = \dfrac{2 \Gamma\sub{tot}} {M_{\rm planet}\Omega a}
\end{equation}
where $\Gamma_{\rm tot}$ is the combined Lindblad and corotation torque exerted by the disc on the planet \citep[e.g.,][]{kley12}:
\begin{equation} \label{gammatot}
    \Gamma\sub{tot} \approx -2.2 \Big( \dfrac{M\sub{planet}}{M_\star} \Big)^2 \Big( \dfrac{a}{H} \Big)^{2} \Sigma\sub{g,planet} a^4 \Omega^2 \,.
\end{equation}
All quantities are evaluated at the location of the planet.
The numerical pre-factor is calibrated using the hydrodynamical
simulations of \citet{dangelo10} for the power-law indices
describing our gas surface density and temperature profiles.
For $M_{\rm planet} > M_{\rm Type\,II}$,
we switch the migration rate from Type I to Type II:
\begin{equation} \label{type2}
    \dot a = -\dfrac{\alpha c\sub{s} H}{a} \times \min (1, \Sigma\sub{g} a^2/M\sub{planet}) \,.
\end{equation}
The correction factor $\min (1, \Sigma\sub{g}a^2/M\sub{planet})$
accounts for how the planet adds an extra load to the disc whose
viscous drift rate is then slowed (note that we assume
$\Sigma\sub{g}$ and not $\Sigma\sub{g,planet}$ is relevant 
for this correction). \citet{duffell14} point out
that the actual Type II rate could differ from
(\ref{type2}) by factors of several 
because disc gas can cross the gap (see also \citealt{Kanagawa18} for an improved treatment of how gap-opening planets migrate); we neglect this effect.

\subsection{Implementation Summary} \label{tech}

There are five input parameters governing the disc: $\alpha$, $a_1$, $s$,
$M\sub{disc,g}(0)$, and $Z \equiv M\sub{disc,s}(0)/M\sub{disc,g}(0)$.
For the planet, we have another three inputs:
the initial core mass $M\sub{core}(0)$, 
the initial gas mass $M\sub{gas}(0)$, 
and the initial orbital radius $a(0)$.
For all simulations, we fix 
$M\sub{core}(0) = 10^{-2} M_\oplus$ 
and $M\sub{gas}(0) =  0$. 
All other parameters are systematically varied
across 540 different models; see Table \ref{tab:var}. 

Equation (\ref{core_eq}), the time derivative of equation
(\ref{dustfree}), and equation (\ref{eq:type1}) (or equation
\ref{type2} once $M_{\rm planet} > M_{\rm Type\,II}$)
define a coupled system of differential
equations for the time evolution of the planet's core mass,
gas mass, and orbital radius.
The background gas disc evolves according to (\ref{sig_g}),
and the background solid disc is evolved according to the
Lagrangian ring scheme described at the end of
section \ref{sdisk}. 
Each model is evolved from $t = 10^{-3}$ Myr to $100$ Myr,
with state variables recorded at 5000 log-spaced times.
To advance from one recording time to the next,
we first evolve our background solid disc using a
fourth-order Runge-Kutta method,
and then solve for the planet variables 
using \texttt{SciPy}'s \texttt{odeint}. 
A core stops growing in solid mass either
once it reaches the pebble isolation mass
(equation \ref{eq:peb_iso})
or the disc of solids drifts wholly interior to the core's orbit. 
Once the planet's gas mass reaches half its core mass,
then either $M_{\rm planet}$ is manually set to 
$1 M_{\rm J}$, or $M_{\rm gas}$ is
augmented by the gas disc mass outside the planet's orbit
at that time, whichever yields the smaller planet mass.
Planets that migrate to our inner
disc radius of $a_{\rm in} = 0.01$ AU are held there.

\section{RESULTS}
\label{sec:results}
\begin{figure}
    \centering
    \includegraphics[width=\columnwidth]{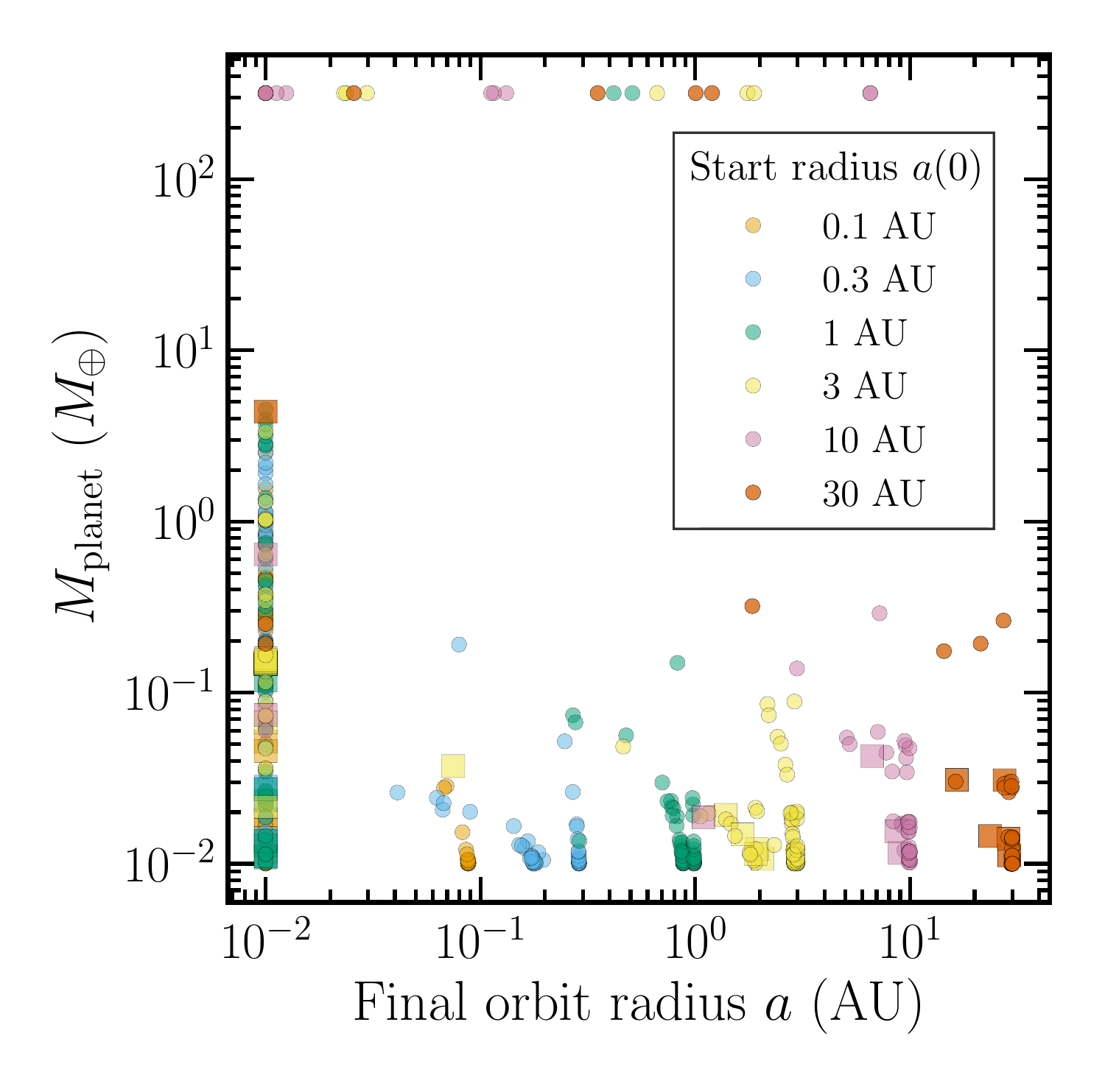}
    \caption{Final planet mass vs.~final orbital distance for 
    each of 540 parameter configurations
    (see Table \ref{tab:var}).
    Each point is coloured according to the initial position of the
    seed core $a(0)$. Those models with a solid drift timescale,
    $a_1 / [v_{\rm drift} (a_1,0)]$, longer than 1 Myr are marked as squares, while discs with shorter drift timescales are marked as circles.
Certain parameter choices lead to
cold, warm, and hot Jupiters (top line of points), all of
which experience migration, and none of which materialize from
long-lived solid discs (no squares, only circles).
No super-Earths/sub-Neptunes form outside of the innermost
disc edge at $0.01$ AU; we are inclined to rule out
those models where super-Earths are found at the innermost
edge at $a = 0.01$ AU having migrated there,
as such models would predict short-period pile-ups in
super-Earth occurrence rates
that are not observed \citep{Lee17}. Over most of our parameter space,
cores hardly grow, remaining $\lesssim 0.1 M_\oplus$. 
}
    \label{fig:a0_scatter}
\end{figure}

\begin{figure*}
    \centering
    \includegraphics[width=\textwidth]{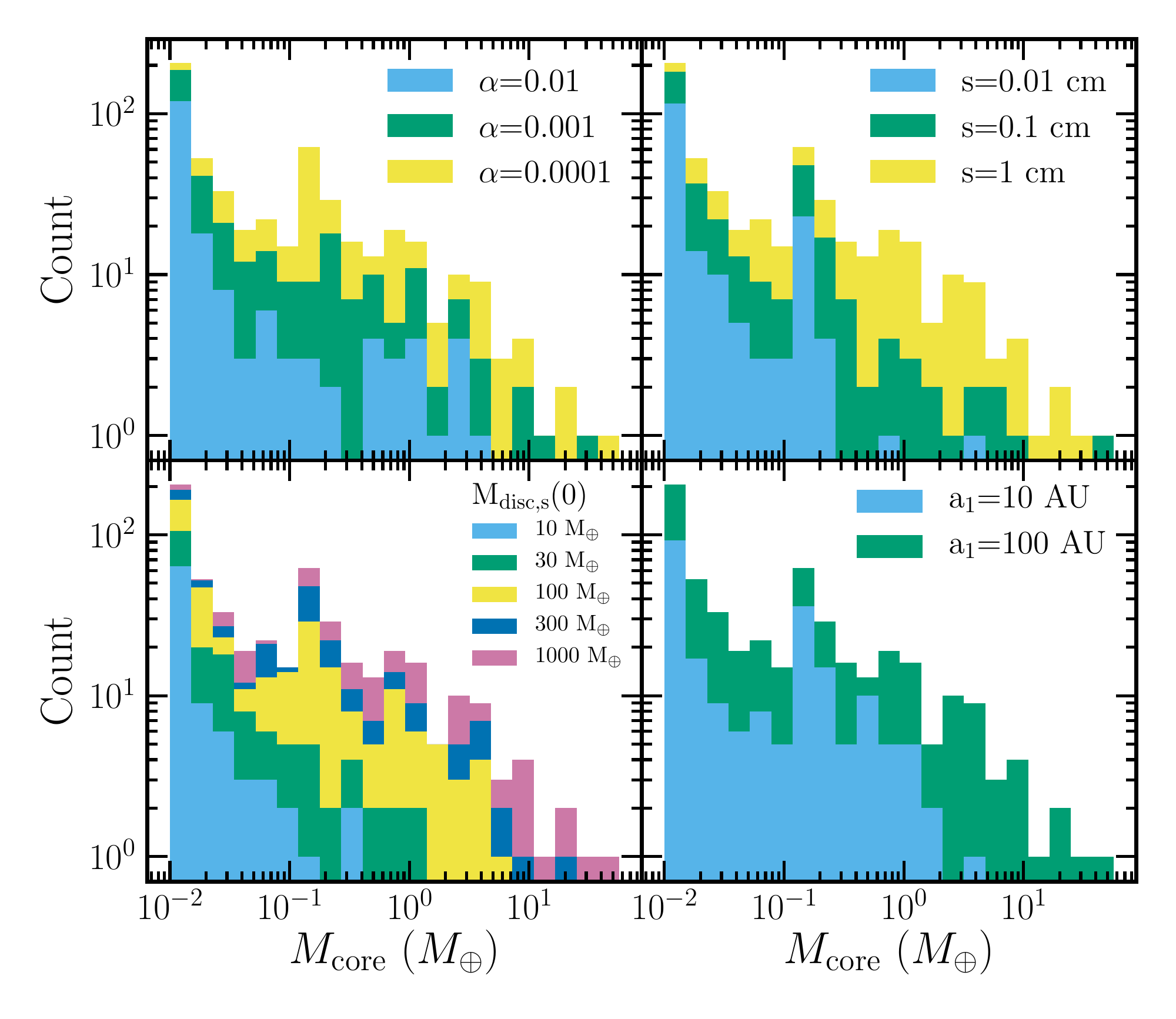}
    \vspace{-5mm}
    \caption{Final core mass histograms, stacked and coloured against various model input parameters. We see that higher mass cores prefer lower $\alpha$, higher $s$, higher ${\rm M}_{\rm disc, s}(0)$, and higher $a_1$. See text for discussion.}
    \label{fig:2x2_hist}
\end{figure*}

\begin{figure}
    \centering
    \includegraphics[width=\columnwidth]{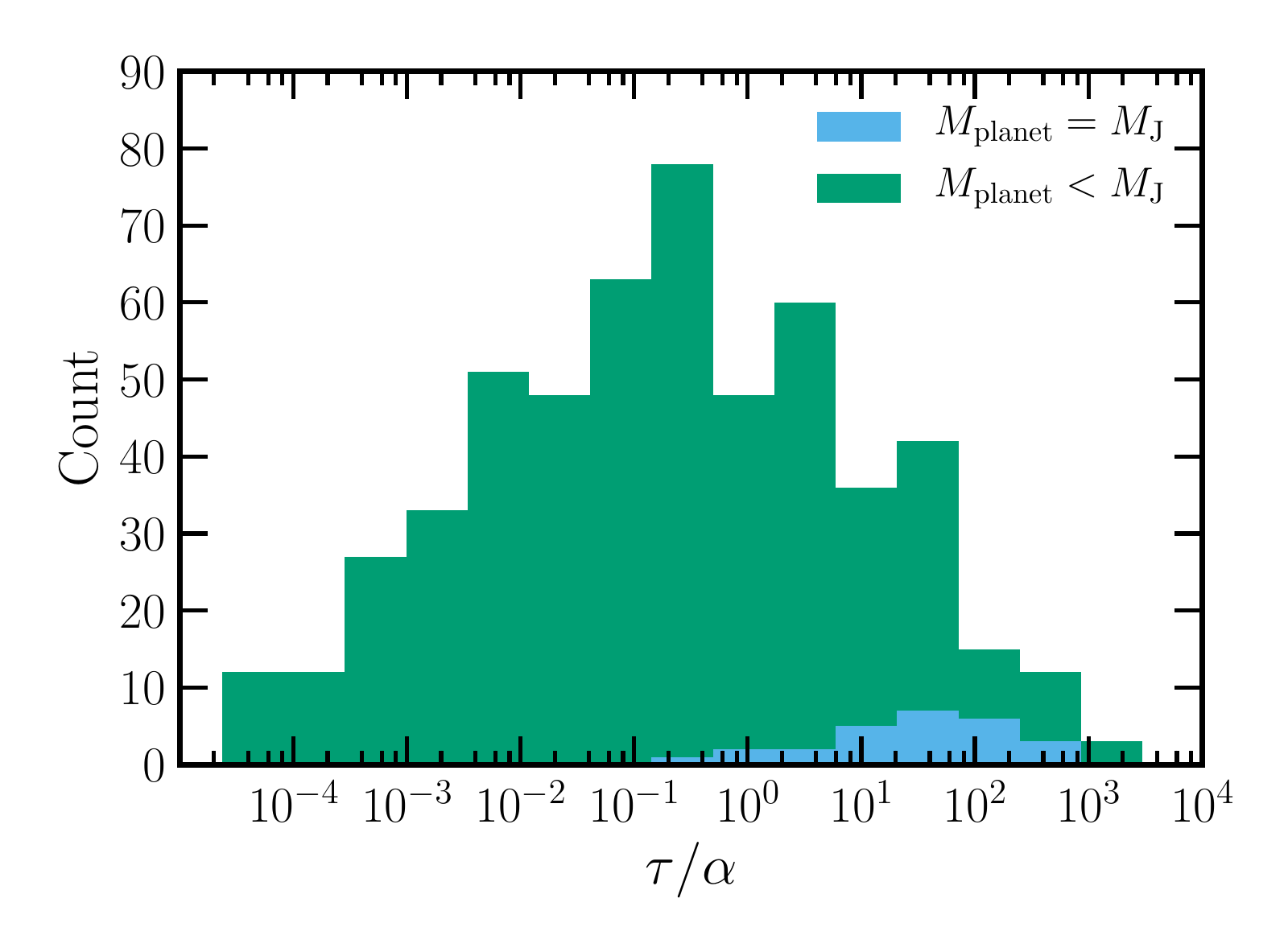}
    \caption{Histogram of the ratio $\tau$/$\alpha$ for our 540 models, where
    Stokes number $\tau$ is evaluated at the initial time (and in practice remains 
    constant over most if not all of the duration of core growth;
    see following figures). The histogram is additionally stacked
    and coloured according to
    whether or not a gas giant emerges from a given model. All gas giants
    arise from model parameter combinations for which $\tau/\alpha > 0.1$.
    This necessary condition on core growth is consistent with analytic calculations; see discussion surrounding equations 
    (\ref{eq:mcore_final_tau_gt_alpha}) and 
    (\ref{eq:mcore_final_tau_lt_alpha}).
    }
    \label{fig:ta_hist}
\end{figure}

\begin{figure}
    \centering
    \includegraphics[width=\columnwidth]{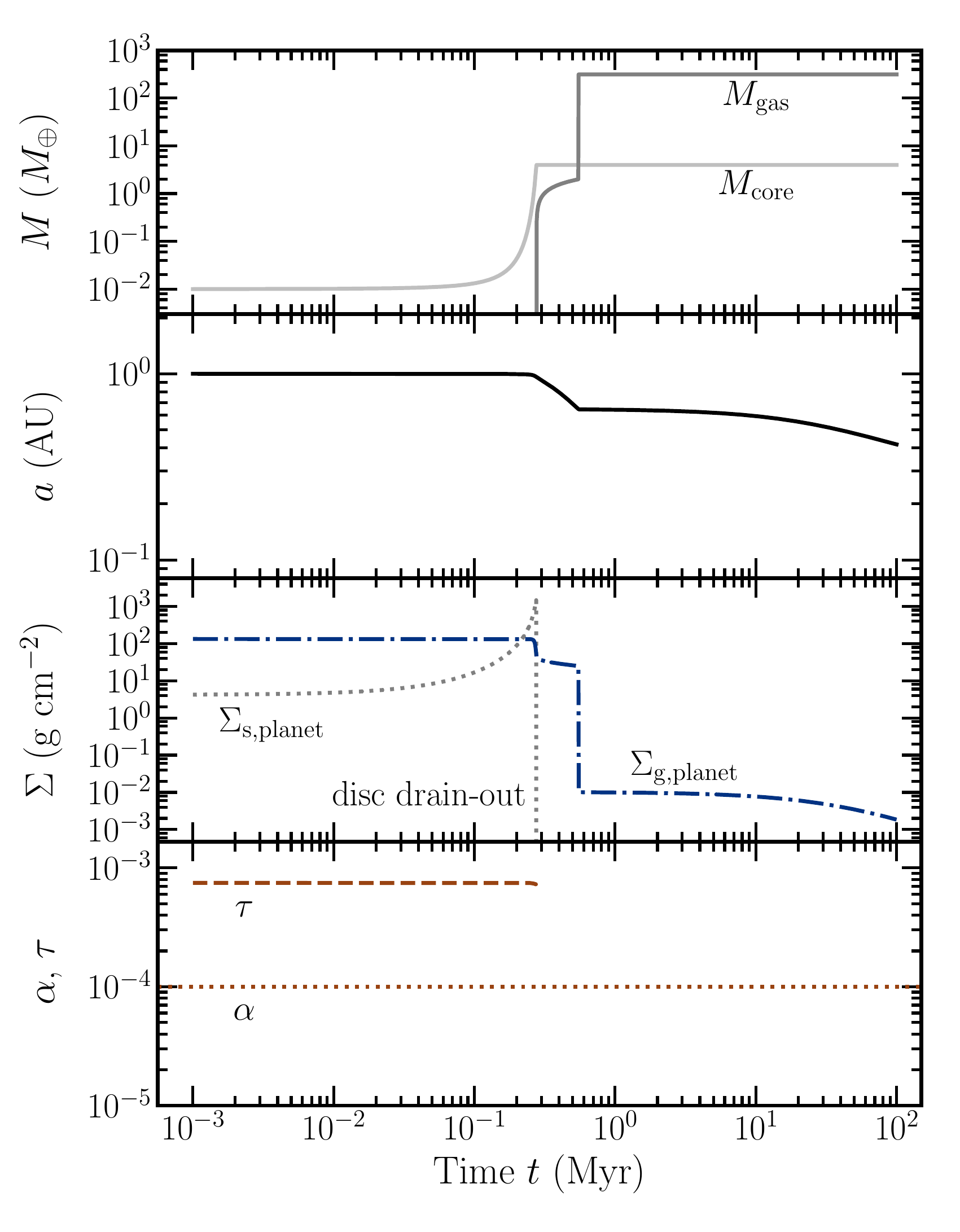}
    \vspace{-5mm}
    \caption{Genesis of a warm Jupiter. All quantities are evaluated at the
    location of the planet. The core mass $M_{\rm core}$
    grows most rapidly when the local solid surface density
    $\Sigma_{\rm s,planet}$ rises from a particle pile-up (compare first
    and third panels from the top). This pile-up just precedes
    the drain-out of solids from outside the core's orbit (see, e.g.,
    Figure \ref{fig:sdisk0}). Dips in $\Sigma_{\rm g,planet}$ reflect the
    deepening of gaps in the gas disc following planet mass growth.
    According to our toy prescription for runaway gas accretion,
    when the planet's gas mass $M_{\rm gas} = 0.5 M_{\rm core}$,
    we instantly set the total planet mass
    $M_{\rm planet} = M_{\rm gas} + M_{\rm core} = 1 M_{\rm J}$.
    The planet has technically not stopped migrating at the end of the
    simulation, but we expect it to eventually park not too far away, given
    the steady decline in $\Sigma_{\rm g,planet}$ from viscous diffusion
    onto the star (the actual migration history will depend on the actual
    dispersal history of the disc which is beyond the scope of this paper). 
    Input parameters for the model shown are: $\alpha=10^{-4}$, $s=0.1$ cm, $a_1=100$ AU, $a(0) = 1$ AU, $M_{\rm disc, g}=10 \,M_{\rm J}$, $M_{\rm disc, s}=100 \,M_\oplus$.
    }
    \label{fig:wj}
\end{figure}

\begin{figure}
    \centering
    \includegraphics[width=\columnwidth]{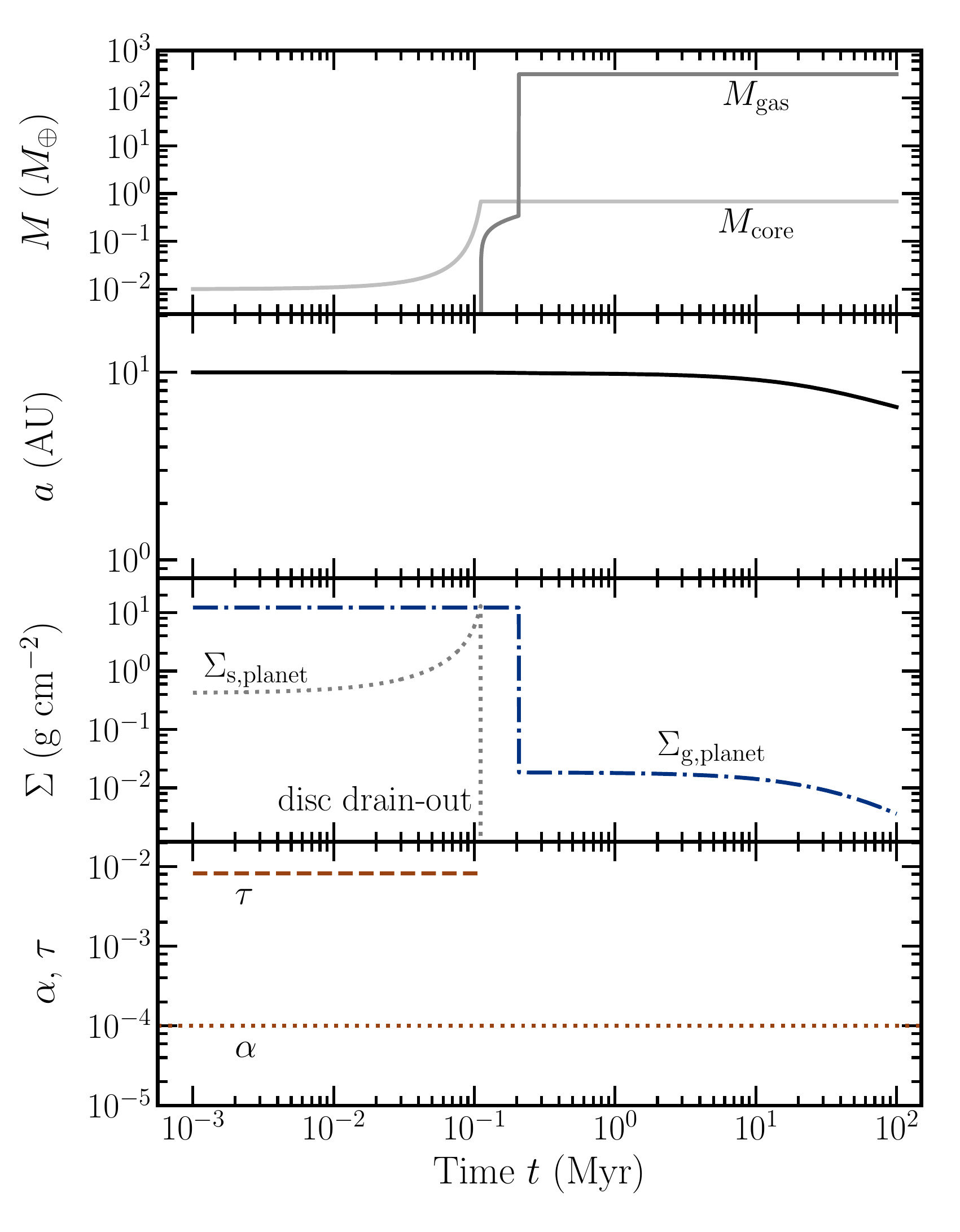}
    \caption{Genesis of a cold Jupiter. Caption text for Figure \ref{fig:wj} applies here. Input parameters: $\alpha=10^{-4}$, $s=0.1$ cm, $a_1=100$ AU, $a(0) =10$ AU, $M_{\rm disc, g}=10 \, M_{\rm J}$, $M_{\rm disc, s}=100 \, M_\oplus$.}
    \label{fig:cj}
\end{figure}

\begin{figure}
    \centering
    \includegraphics[width=\columnwidth]{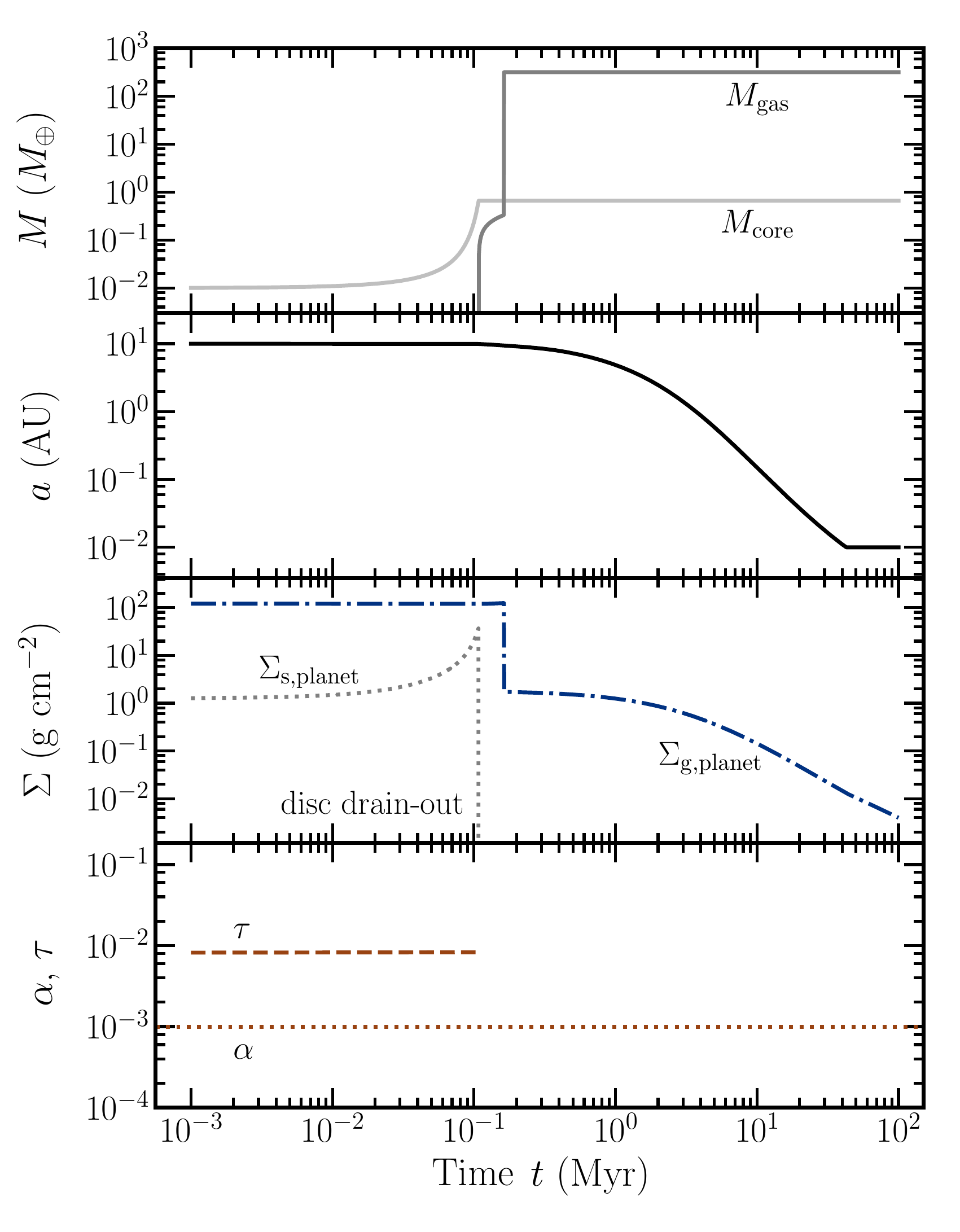}
    \caption{Genesis of a hot Jupiter. Caption text for Figure \ref{fig:wj}
    applies here. Input parameters: $\alpha=10^{-3}$, $s=1$ cm, $a_1=100$ AU, $a(0)=10$ AU, $M_{\rm disc, g}=100 \, M_{\rm J}$, $M_{\rm disc, s}=300 \, M_\oplus$.}
    \label{fig:hj}
\end{figure}

\begin{figure}
    \centering
    \includegraphics[width=\columnwidth]{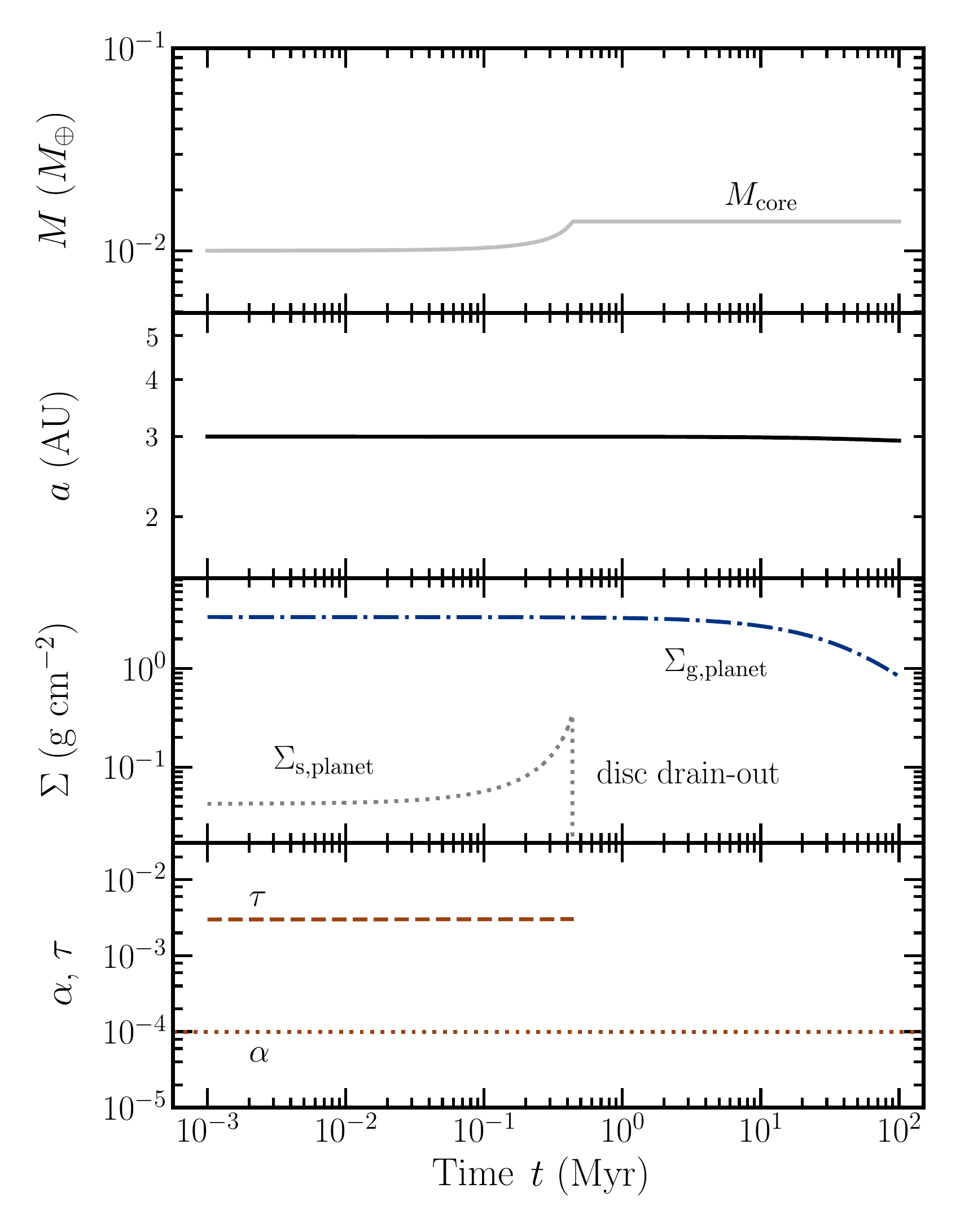}
    \caption{A case where pebble accretion fails to grow much
      of anything. Although $\tau > \alpha$ (see Figure \ref{fig:ta_hist}),
      the total inventory of solids $M_{\rm
        disc,s}$ is too small. Plot format follows that of Figure
      \ref{fig:wj}. Input parameters: $\alpha=10^{-4}$, $s=0.01$ cm,
      $a_1=100$ AU, $a(0)=30$ AU, $M_{\rm disc,g}=10 \, M_{\rm J}$,
      $M_{\rm disc,s}=30 \, M_\oplus$.}
    \label{fig:fc2}
\end{figure}

\begin{figure}
    \centering
    \includegraphics[width=\columnwidth]{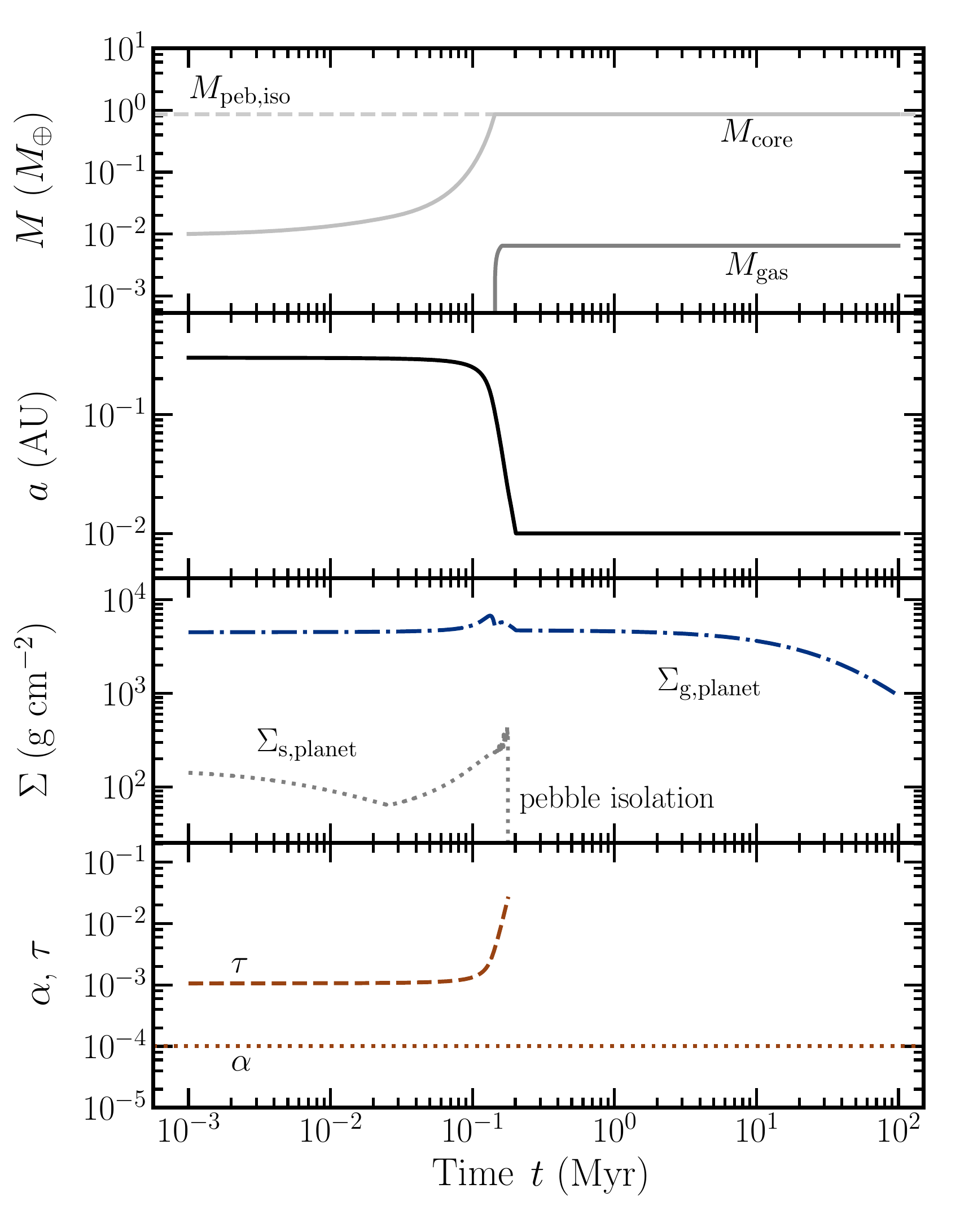}
    \caption{A case where pebble accretion forms a hot Earth but not a
      Jupiter. Although $\tau>\alpha$ and $M_{\rm disc, s}$ is
      assigned its highest possible value, a gas giant fails to form
      because the planet's orbital radius starts small and only gets
      smaller by migration. Small orbits are more susceptible to
      pebble isolation ($M_{\rm peb,iso} \propto a^{3/4}$; see
      equations \ref{eq:peb_iso} and \ref{eq:H}), which is what
      ultimately limits the core mass here.
      Gas accretion is quickly terminated once the atmospheric mass reaches the upper bound appropriate to an isothermal 
      atmosphere---an upper bound 
      made low by the high temperature ($2600$ K) at the innermost disc
      edge. Plot format follows that of Figure
      \ref{fig:wj}.  Input parameters: $\alpha=10^{-4}$, $s=1$ cm,
      $a_1=100$ AU, $a(0)=0.3$ AU, $M_{\rm disc, g}=100 \, M_{\rm J}$,
      $M_{\rm disc, s}=1000 \, M_\oplus$.}
    \label{fig:fc3}
\end{figure}

\begin{figure}
    \centering
    \includegraphics[width=\columnwidth]{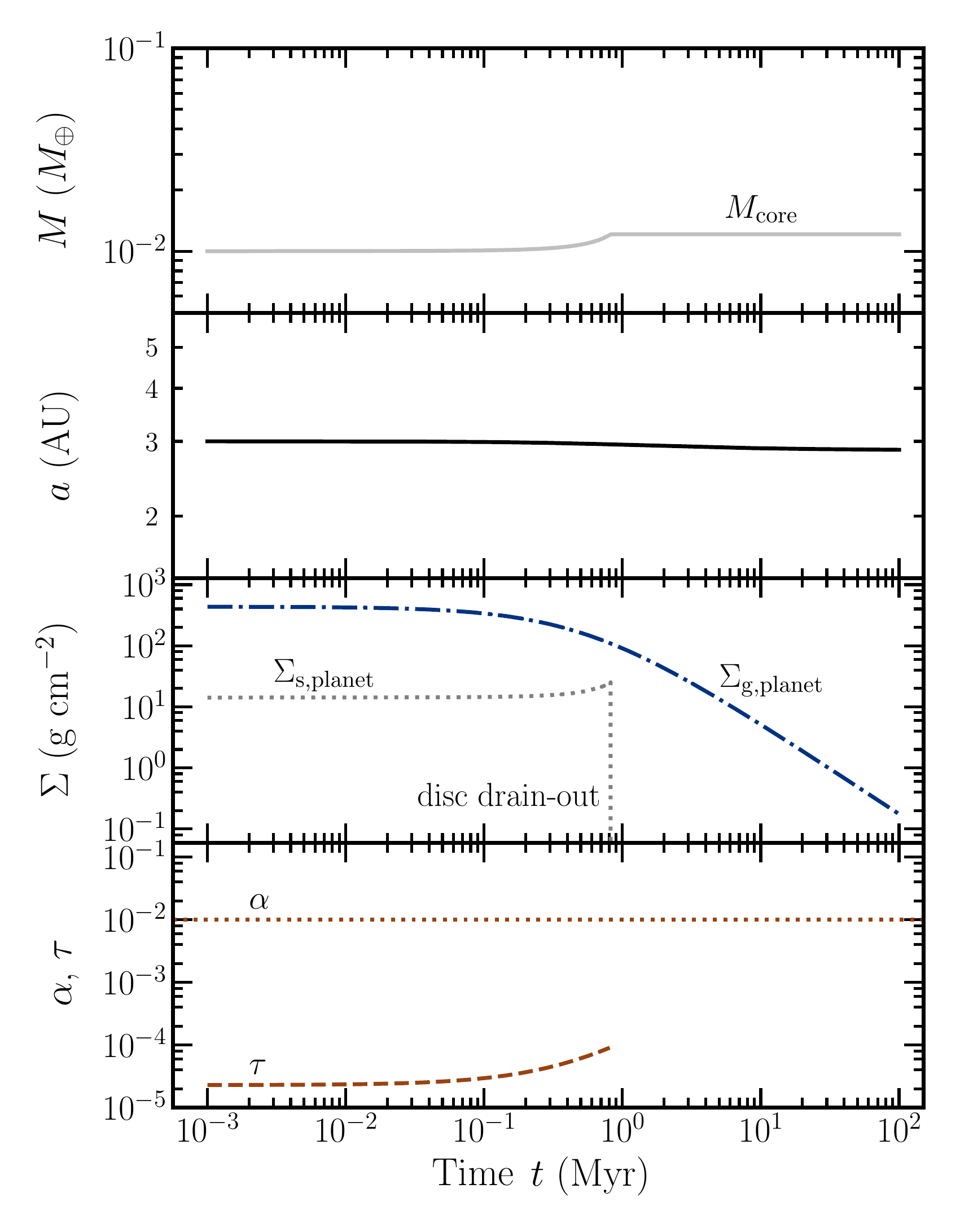}
    \caption{A case where pebble accretion is rendered practically
      impotent, here because $\tau \ll \alpha$; particles are too
      strongly coupled to gas to drop out onto cores, and particle
      densities are made too dilute by turbulent stirring. Furthermore, 
      because $\tau \ll \alpha$, the particle pile-up at the planet's
      position is more muted than in other figures in this series.
      Plot format follows that of Figure \ref{fig:wj}. Input parameters:
      $\alpha=10^{-2}$, $s=0.01$ cm, $a_1=100$ AU, $a(0)=3$ AU,
      $M_{\rm disc, g}=100 \, M_{\rm J}$,
      $M_{\rm disc, s}=1000\, M_\oplus$.}
    \label{fig:fc4}
\end{figure}

Figure \ref{fig:a0_scatter} 
summarizes our model outcomes.
Beyond $\sim$0.1 AU---in those regions where the majority
of exoplanets are detected---we find that sub-Earth cores
either remain sub-Earths,
or explode into gas giants. There is no in-between;
super-Earths are completely absent outside the assumed disc
edge at 0.01 AU.
Over most of the parameter space that we chart,
sub-Earths are the norm: they outnumber gas giants 
by about 20 to 1.
Moreover, these sub-Earths have hardly grown from their
initial assumed seed
masses of $10^{-2} M_\oplus$; on the whole, they have increased
their mass by factors of several at most.

A small fraction (6\%) of models produce super-Earths 
($\sim$1--10$M_\oplus$).
Formed overwhelmingly at short distances ($a(0) \leq 1$ AU) 
and/or in gas-heavy discs, 
the cores of these super-Earths rapidly migrate 
and become stranded at the disc innermost edge,\footnote{When our standard 
prescription for fast dust-free gas accretion
is replaced with a slower one using opacities that include dust
(footnote \ref{foot:predusty}),
those models that previously formed gas giants mostly
form super-Earths that migrate and pile up at the innermost disc edge.
The remainder continue to form gas giants
with no substantive change in outcome (this is
the case when solid disc masses and initial core distances
are large).\label{foot:dusty}}
where the surrounding nebula is so hot that 
gas accretion onto cores is stunted---the atmospheric masses
hit their isothermal upper bounds (evaluated at
$T = 2600$ K at $a = 0.01$ AU).
Final gas-to-core mass ratios are about 1--10\%,
with a few reaching up to 30\%.
These migration-heavy
models predict that planets pile up at short orbital periods
\citep[see, e.g.,][]{ida08}.
Because such pile-ups
are not seen in observations \citep[see, e.g.,][and references 
therein]{Lee17},
we tend to discount as unrealistic those model parameters
that lead to such wholesale migration of cores.

Whether a core nucleates a gas giant depends on its mass: more massive
cores accrete gas faster \citep{pollack96,piso14,Lee15,Ginzburg16}.
In the following,
we describe analytically the factors that determine how quickly and to what
final mass a core grows by pebble accretion---including how much disc mass
is expended in the process. These pencil-and-paper considerations
help to explain our numerical results, which are also fleshed out
in greater detail below.

The final mass of the core
is determined by how many pebbles the seed core 
nets from the background drift
of solids (orbital migration of the core can be safely
neglected during this early growth phase).
From equation (\ref{vdrift}) for $v_{\rm drift}$,
the time it takes for the solid disc
to drain from its initial outer radius
$a_1$ to the location of the core $a$ is
\begin{equation}
\label{eq:tdrift}
    t_{\rm drift} \simeq
    \begin{cases}
            a/(2v_{\rm hw}\tau) \ln{(a_1/a)}, & \tau \gg \alpha \\
          (2/3) \alpha^{-1} (\Omega a/c_{\rm s}^2) (a_1-a), & \tau \ll \alpha
    \end{cases}
\end{equation}
where all unsubscripted variables (here and below)
are evaluated at the position
of the core. We have taken the limit $\tau \ll 1$ (valid over practically
all of our parameter space) and made use of the fact that 
for our assumed disc gas surface density and temperature  profiles, $\tau \propto a$
and $v_{\rm hw} =$ constant.

Because pebble accretion for our parameters occurs mostly
in the settling regime ($\tau < 1$), and with $R_{\rm acc} < H_{\rm s}$
(see Appendix \ref{app:eff} for a more general exposition; see also
\citealt{Ormel2017}), we can write:
\begin{align} 
    \dot{M}_{\rm core} &= 2 \Sigma_{\rm s} R_{\rm acc}^2 v_{\rm acc} / H_{\rm s} \nonumber \\
    &= \frac{8\Sigma_{\rm s}GM_{\rm core}\tau}{c_{\rm s}}\left(\frac{\alpha + \tau}{\alpha}\right)^{1/2} \,. \label{eq:mdot_same}
\end{align}
Accordingly, the core grows exponentially with time:
\begin{equation}
\label{eq:mcore}
    M_{\rm core}(t) = M_{\rm core}(0)\,{\rm exp}\left[\frac{8G\tau}{c_{\rm s}}\left(\frac{\alpha + \tau}{\alpha}\right)^{1/2}\int^t_0\Sigma_{\rm s}dt\right] 
\end{equation}
assuming $\tau$ at the position of the core is constant with time (which
it approximately is in our models at early times; see our later figures). 
For $\tau \gg \alpha$, we can use the relations
$v_{\rm drift} \propto v_{\rm hw} \tau \propto a$ (Epstein) 
and $\Sigma_{\rm s} \propto a^{-1}$ to solve the continuity
equation for $\Sigma_{\rm s}$ by separation of variables:
$\Sigma_{\rm s}(a,t)= f(a) g(t) \propto a^{-1} g(t)$, whence
$g(t) = \exp [t/(a/v_{\rm drift})]$ (i.e., an exponentially rising ``particle
pile-up'' at fixed location; see, e.g., Figure \ref{fig:sdisk0}).
Then the final core mass after $t = t_{\rm drift}$ is
\begin{align}
\label{eq:mcore_final_tau_gt_alpha}
    M_{\rm core} \sim M_{\rm core}(0)\,{\rm exp}\left[\frac{2}{\pi}\left(\frac{\tau}{\alpha}\right)^{1/2}\frac{M_{\rm disc,s}(0)}{M_\star}\left(\frac{\Omega a}{c_{\rm s}}\right)\left(\frac{\Omega a}{v_{\rm hw}}\right)\right] \,,\nonumber \\
    \tau \gg \alpha
\end{align}
where we have used $M_{\rm disc,s}(0) = 2\pi \Sigma_{\rm s}(a,t=0) a a_1$.
For $\tau \ll \alpha$,
$\Sigma_{\rm s}$ is approximately constant in time
(since $\Sigma_{\rm s} \propto a^{-1}$ is the steady-state solution
for $v_{\rm drift} = 3\nu/(2a) =$ constant), and so 
\begin{align}
\label{eq:mcore_final_tau_lt_alpha}
    M_{\rm core} \sim M_{\rm core}(0)\,{\rm exp}\left[\frac{8}{3\pi}\left(\frac{\tau}{\alpha}\right)\frac{M_{\rm disc,s}(0)}{M_\star}
    \left(\frac{\Omega a}{c_{\rm s}}\right)^3
    \frac{a_1 - a}{a_1}
    \right] \,, \nonumber \\
    \tau \ll \alpha \,.
\end{align}
From equations (\ref{eq:mcore_final_tau_gt_alpha}) 
and (\ref{eq:mcore_final_tau_lt_alpha}), it follows that more massive
cores, and by extension Jupiters, favour high $M_{\rm disc,s}(0)$, 
high $\tau$ (to enable particles to more
easily ``peel off'' the gas flow and fall onto the core),
and low $\alpha$ (to reduce the particle scale height and increase
the solid particle density; and, in cases where the pebble drift speed
is determined by viscosity, to slow that drift speed down and prolong
the time over which the core accretes).
Figure \ref{fig:2x2_hist} verifies these dependencies, at least in sign.
More massive cores are formed in discs with higher
overall solid mass (lower left panel), smaller $\alpha$
(upper left panel), larger pebble size $s$
(which increases $\tau$; upper right panel and equation \ref{tau}),
and larger $a_1$ (which also increases $\tau$ by spreading a given
gas mass over a larger area to reduce the gas density; lower
right panel and equation \ref{tau}).

Equations (\ref{eq:mcore_final_tau_gt_alpha}) and
(\ref{eq:mcore_final_tau_lt_alpha}) further
suggest that the ratio $\tau/\alpha$ is a key parameter controlling
how massive cores can grow. 
Figure \ref{fig:ta_hist} bears this out by showing
that Jupiter-breeding cores only form when $\tau/\alpha > 0.1$.
The condition $\tau/\alpha > 0.1$ 
is necessary but not sufficient
to create gas giants;
the vast majority of our runs at $\tau/\alpha > 0.1$
yield only sub-Earths, mostly because $M_{\rm disc,s}(0)$ is
too small. Another necessary condition for spawning giants
is that their cores grow 
outside $\sim$0.1 AU---sufficiently far away from their host stars that 
they do not run up against the pebble isolation mass, which decreases
with decreasing distance to the star ($M_{\rm peb,iso} \propto a^{3/4}$;
see equations \ref{eq:peb_iso} and \ref{eq:H}).
This limitation set by pebble isolation 
argues 
against in-situ formation of hot Jupiters. 
An example of core growth stifled by pebble isolation 
will be given below.

A core accretes only a fraction of the solids from the outer disc 
that converge onto its orbit. The remaining fraction drifts past the core
into the inner disc and is ``wasted''.
To grow a core from an initial seed mass $M_{\rm core}(0)$
requires a mass investment of
\begin{equation}
\label{eq:Mdrift}
    M_{\rm drift} = \int^{M_{\rm core}}_{M_{\rm core}(0)}\epsilon^{-1} dM_{\rm core}
\end{equation}
where the instantaneous pebble accretion efficiency
\begin{align}
\label{eq:epsilon}
    \epsilon &\equiv \frac{\dot{M}_{\rm core}}{2\pi\Sigma_{\rm s}v_{\rm drift}a} \nonumber \\
    &= \frac{4}{\pi} \left(\frac{\Omega a}{v_{\rm drift}}\right)\left(\frac{\Omega a}{c_{\rm s}}\right)\left(\frac{M_{\rm core}}{M_\star}\right)\tau \left(\frac{\alpha + \tau}{\alpha}\right)^{1/2} \,.
\end{align}
Note how $\epsilon$, and by extension $M_{\rm drift}$,
do not depend on $\Sigma_{\rm s}$ and the details of its time evolution.
Evaluation gives:
\begin{align}
\label{eq:Mdrift_specific}
&M_{\rm drift} = \frac{\pi}{4} \left(\frac{v_{\rm drift}}{\Omega a}\right)\left(\frac{c_{\rm s}}{\Omega a}\right)\tau^{-1} \left(\frac{\alpha}{\alpha+\tau}\right)^{1/2}M_\star\ln{\left[\frac{M_{\rm core}}{M_{\rm core}(0)}\right]} \nonumber \\
    &\sim \begin{cases}
          35\,M_\oplus\left(\frac{a}{1\,{\rm AU}}\right)^{3/4}\left(\frac{10}{\tau/\alpha}\right)^{1/2}\ln{\left(\frac{M_{\rm core}/M_{\rm core}(0)}{100}\right)}, &\tau \gg \alpha \\
          600\,M_\oplus\left(\frac{a}{1\,{\rm AU}}\right)^{3/4}\left(\frac{0.1}{\tau/\alpha}\right)\ln{\left(\frac{M_{\rm core}/M_{\rm core}(0)}{100}\right)}, &\tau \ll \alpha \,.
    \end{cases}
\end{align}
These analytic estimates appear consistent with our numerical results
as reported in Figure \ref{fig:2x2_hist} (lower left panel). 

Example evolutionary
tracks of seed cores that grow to
gas giants are presented in Figures \ref{fig:wj}
(warm Jupiter), \ref{fig:cj} (cold Jupiter),
and \ref{fig:hj} (hot Jupiter).
For comparison we also show simulations that terminate in
sub-Earths in Figures
\ref{fig:fc2}, \ref{fig:fc3}, and \ref{fig:fc4}, illustrating
three different modes by which gas giant formation by pebble accretion
can fail (low $M_{\rm disc,s}(0)$; low $M_{\rm peb,iso}$;
and $\tau/\alpha < 0.1$, respectively).

For all evolutionary tracks, we verify that the core mass grows
at least exponentially fast.
Note how in those runs for which $\tau > \alpha$
(Figures \ref{fig:wj}--\ref{fig:fc3}), the solid
surface density $\Sigma_{\rm s}$
at the location of the core rises with time just
before the solid disc drifts entirely past the core.
This ``particle pile-up''---a traffic jam in disc solids---occurs 
whenever the particle drift velocity $v_{\rm drift}$ decreases 
sufficiently fast with decreasing radius \citep{youdin04}. We have verified that
such an inwardly decreasing velocity
profile obtains when $\tau/\alpha > 0.1$---so that 
the drift velocity in (\ref{vdrift}) 
is not dominated by viscous diffusion but has a significant
contribution from aerodynamic drag---and when
that drag is in the Epstein regime for $\tau \ll 1$.
The latter drag regime typically holds  
between $\sim$0.1 and $\sim$10 AU for our model parameters; 
outside $\sim$10 AU,
$\tau$ approaches unity and the velocity profile flattens,
while inside $\sim$0.1 AU, Stokes drag obtains which bleeds
particles from the inside out. Where there is a particle pile-up,
core growth is super-exponentially fast (equation \ref{eq:mcore}).

Although we have identified model parameters 
that succeed in forming Jupiters by pebble accretion, 
these same parameters encounter difficulty when 
confronted with millimetre-wavelength observations of discs. 
The problem is that rapid core growth by pebble
accretion demands large $\tau$ (equations 
\ref{eq:mcore_final_tau_gt_alpha} and \ref{eq:mcore_final_tau_lt_alpha}),
but that same large $\tau$
leads to solids draining quickly from the disc
(equation \ref{eq:tdrift})---too quickly
when compared against observations. 
In all of our Jupiter-forming runs, solids
drain out in $\sim$0.1--0.3 Myr
(Figures \ref{fig:a0_scatter} and \ref{fig:wj}--\ref{fig:hj})
or shorter (data not shown). 
These results cannot be immediately reconciled with
observed discs that orbit stars 1--10 Myr old and that exhibit
mm-wave continuum emission---presumably from mm-sized solids---on
scales of 10--100 AU  \citep[e.g.,][]{Brauer07,Tripathi17,Tazzari16,Perez15}.

\begin{figure}
    \centering
    \includegraphics[width=\columnwidth]{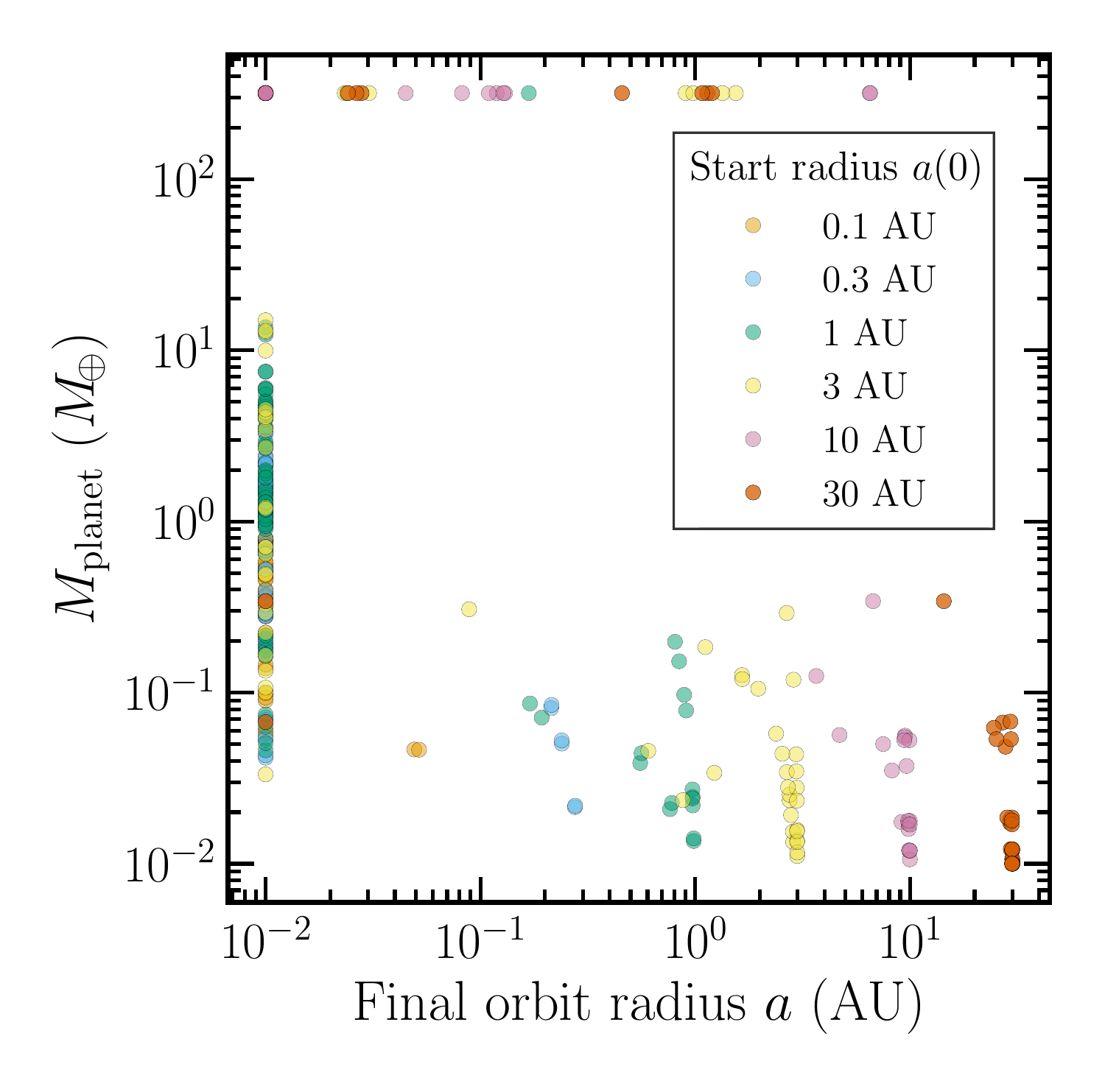}
    \caption{Same as Figure \ref{fig:a0_scatter}, but for
    fixed Stokes number $\tau \in \{0.01, 0.1\}$, in lieu
    of fixed particle size
    $s \in \{ 0.01, 0.1, 1\}$ cm.
    All models shown have a solid drift timescale,
    $a_1 / [v_{\rm drift} (a_1,0)]$, shorter than 1 Myr
    (i.e., there
    are no squares, only circles, unlike in
    Figure \ref{fig:a0_scatter}).
}
    \label{fig:a0_scatter_fixtau}
\end{figure}

\begin{figure}
    \centering
    \includegraphics[width=\columnwidth]{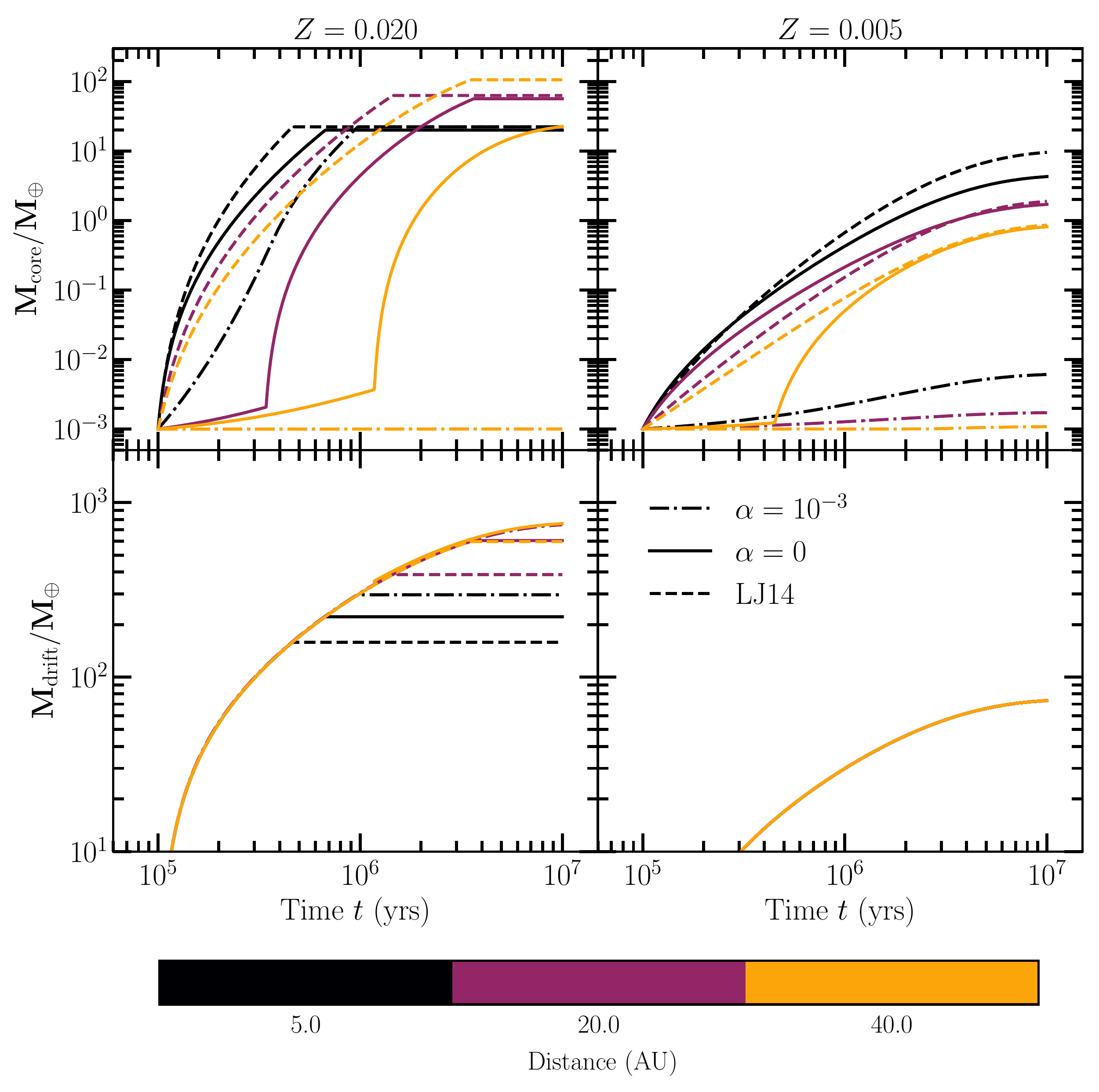}
    \vspace{-5mm}
    \caption{Comparison between the pebble accretion solution of
    \citet[][LJ14, dashed]{Lambrechts14} and solutions that
    explicitly account for $\alpha$.
    To isolate the dependence on $\alpha$, 
    we use the same input parameters as LJ14---in particular, we incorporate
    their grain growth prescriptions, as outlined in section 2 of LJ14 (their
    equations 20 and 25); their initial core mass of $10^{-3} M_\oplus$; 
    and their fiducial gas surface density
    $\Sigma_{\rm g} = 500\,{\rm g\,cm^{-2}}(a/{\rm AU})^{-1}\exp{(-t/3\,{\rm Myr})}$.
    For $\alpha = 10^{-3}$ (dot-dashed), accretion is not 2D 
    ($R_{\rm acc} < H_{\rm s}$) and is
    therefore less efficient than in the
    LJ14 solution. For $\alpha = 0$ (solid),
    accretion is 2D, as was
    assumed by LJ14, but accretion is not always 
    in the settling regime, 
    contrary to the assumption of LJ14. 
    {\it Top:} Core mass vs.~time. Core growth is truncated once the pebble
    isolation mass is reached (20$M_\oplus(a/5{\rm AU})^{3/4}$, equation 34 of LJ14). The assumption of shear-dominated, 
    settling 
    accretion is seen to overestimate
    core growth rates at large distances and early times.
    At 40 AU, cores hardly grow for $\sim$1 Myr, especially for a disc-integrated, initial
    solid-to-gas mass ratio $Z$ ($=Z_0$ in the notation of LJ14) of 0.02.
    In the grain growth model of LJ14, the higher $Z$ is, the more rapidly
    particles grow, and the larger is $\tau$; for $Z = 0.02$,
    the stopping time is longer than the core-particle encounter time
    at $a = 40$ AU for the first $\sim$1 Myr, and pebble accretion is in the (slow) hyperbolic
    regime. {\it Bottom:} The cumulative
    ``wasted'' mass that drifts past the core as the core is growing.
    All model curves necessarily overlap except when the core in a given model stops growing. Comparing our solutions in the top
    panel with our solutions in the bottom panel,
    we infer net pebble
    accretion efficiencies that are at most $M_{\rm core}/M_{\rm drift} \simeq 1/10$ ($Z = 0.02$, $a = 5$ AU, $\alpha = 0$).}
    \label{fig:Mcore_v_t_compLJ14}
\end{figure}

\begin{figure}
    \centering
    \includegraphics[width=\columnwidth]{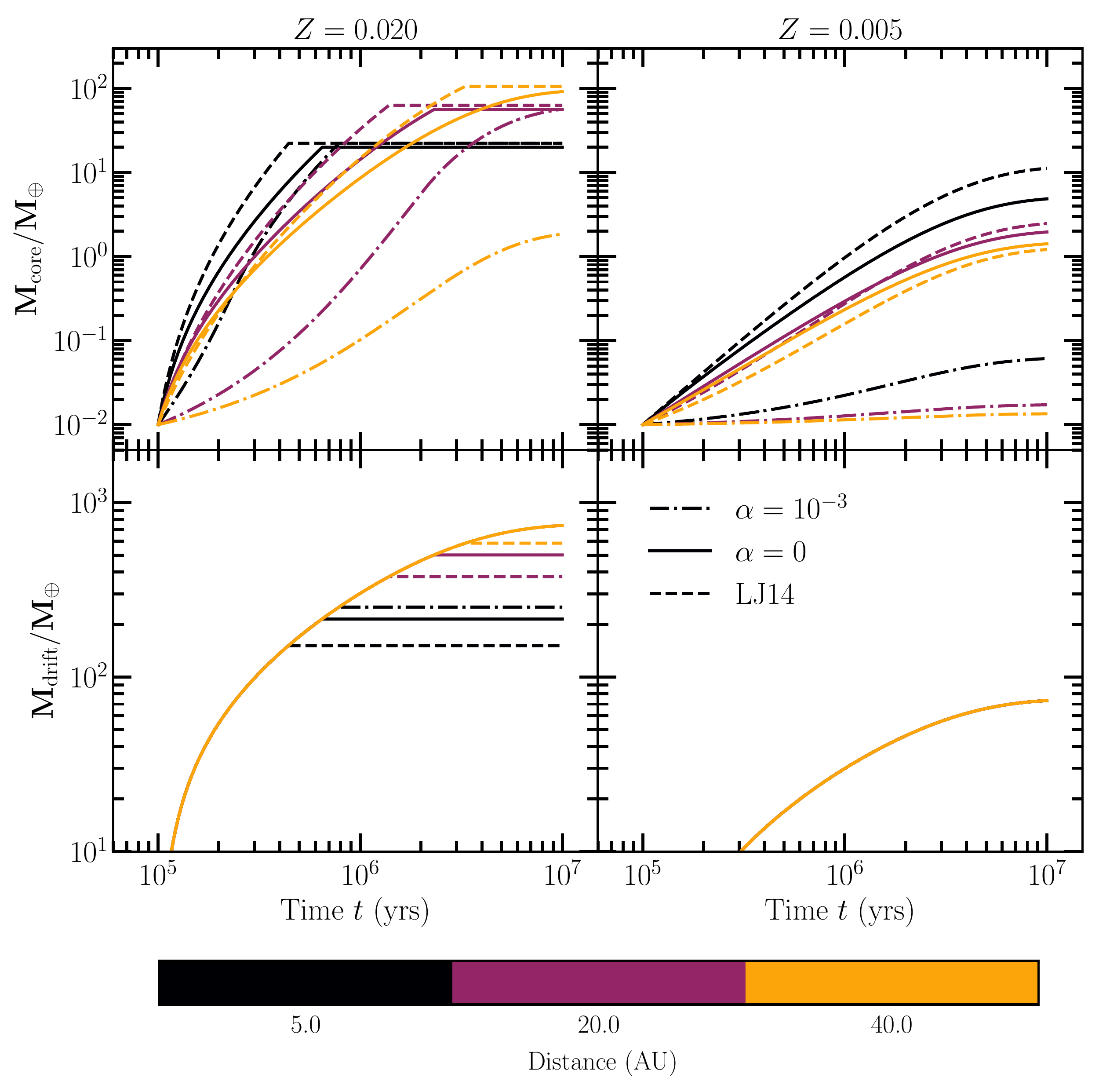}
    \caption{Same as Figure \ref{fig:Mcore_v_t_compLJ14} but starting with a larger seed core of 0.01$M_\oplus$, the mass at which 
    pebble accretion might begin in earnest \citep[see][their Figure 8]{Johansen17}. With a larger starting mass, there is 
    better agreement between the explicit $\alpha$ models
    and LJ14, although at $\alpha - 10^{-3}$
    growth is still slow at $a = 40$ AU.
}
    \label{fig:Mcore_v_t_compLJ14_lunar}
\end{figure}

\begin{figure}
    \centering
    \includegraphics[width=\columnwidth]{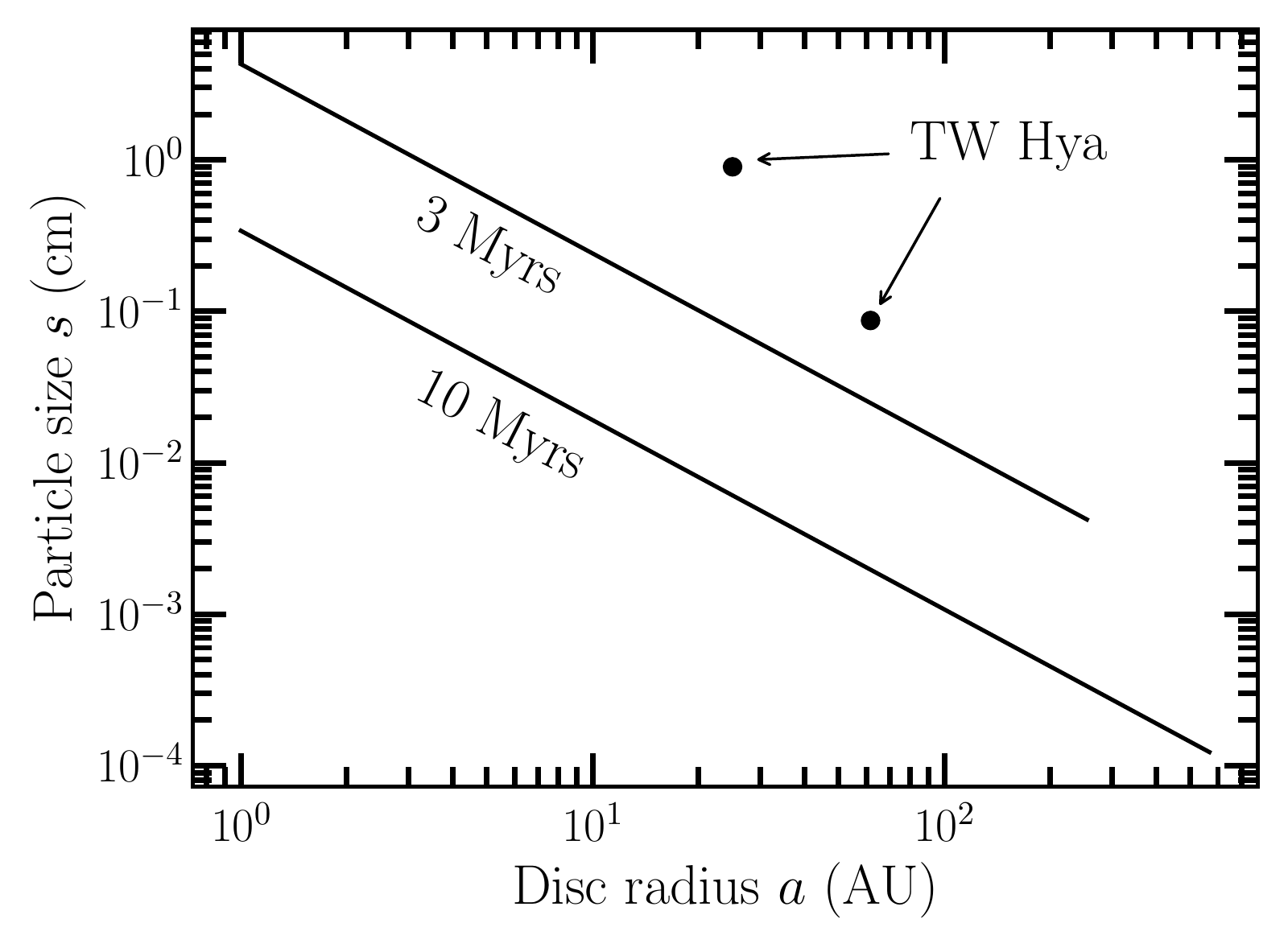}
    \vspace{-5mm}
        \caption{Comparing the grain growth model of 
    \citet[][LJ14, solid lines]{Lambrechts14} to the observed mm--cm
    continuum sizes of the 3--10 Myr old TW Hydra disc
    (\citealt{andrews12,menu14}; black circles).
    The latter data are placed on this plot by assuming that the size
    of the emitting particles equals the wavelength of
    observation. The fiducial model of LJ14 predicts insufficiently
    large particles at 20--100 AU at an age of 3--10 Myr; by this
    time, mm--cm sized objects have drained out by gas
    drag. Staying within
    the framework of the LJ14 grain growth model, this discrepancy
    can be resolved by assuming a longer-lived gas disc (one with
    an e-folding time of 10 Myr),
    and a disc-integrated dust-to-gas ratio
    that is strongly supersolar ($Z = 0.1$) to enhance grain growth.
    An alternative (still within the context of the LJ14 model) is
    to make the disc so gas-heavy (approximately
    10 times more massive than
    their fiducial disc) that it is Toomre $Q$-unstable
    ($Q \sim 0.4$ at $a = 25$ AU). 
     We note in passing that using the model disc of \citet{Bitsch15a, Bitsch15b}, which is approximately 3 times more dense than the fiducial LJ14 disc beyond $\sim$10 AU, produces model curves that approach
    but still do not match the observed data for TW Hya.}
    \label{fig:grain_growth}
\end{figure}

\section{Summary and Discussion}\label{sec:sum}

We have studied how planets can form by pebble accretion, starting
from an assumed seed mass of $10^{-2} M_\oplus$ and working our way to
cores massive enough to nucleate gas giants.  Our model is global in
the sense that it accounts for the parent disc over its entire radial
extent. We calculated how solids drift from large to small orbital
radius by 
aerodynamic drag within viscously spreading gas, and how the disc has
only a finite reservoir of solids with which to build planets.  Prescriptions for pebble
accretion were taken from \citet{Ormel10} and modified for gas
turbulence, while those for gas accretion onto cores were drawn from
\citet{Lee15,Lee16}.

A fixed pebble size of 0.01--1 cm was assumed,
motivated by millimeter-wave disc observations that probe these
very size particles, and by order-of-magnitude considerations
of the limit to which particles can grow by sticking
(e.g., \citealt{Chiang10}, their section 4). In what follows,
we will present some auxiliary calculations that relax this
assumption and utilize more sophisticated
grain growth prescriptions.

We summarize our results as follows, placing them into context
with observations:
\begin{enumerate}
\item {\em Growth by pebble accretion is exponentially sensitive to
solid disc mass.} The solid disc mass controls not only how massive a core
can grow, but also how fast it grows. That growth is at least
exponentially fast during the earliest stages if not the entire duration 
of pebble accretion, with an e-folding time that scales inversely with
the disc solid surface density. Growth can be super-exponentially fast
if outer disc solids ``pile up'' at the position of the core
as they drift inward.

This strong sensitivity of core growth to disc solid content 
(see, e.g., \citealt{Bitsch15a,Bitsch15b} and \citealt{Lambrechts14} for the same qualitative point),
coupled with the need for Jupiters to nucleate from sufficiently
massive cores, accords with the observation that the occurrence 
rates of gas giants \citep{fischer05} and of larger planets
more generally \citep{Buchhave14,petigura18} correlate with host star metallicity (to the extent that the latter
can be used as a proxy for solid disc content). Although the expectation
that discs with more solids spawn more massive cores may seem obvious,
and is not specific to pebble accretion but also characterizes core
formation by giant impacts \citep{dawson15}, it is not a
universal prediction of theory---not even in the context of pebble
accretion.  For example, if the final core mass were limited instead by the
pebble isolation mass \citep[e.g.,][]{Lambrechts+14}, a
correlation between planet radius and host star metallicity
would not be expected, as the
pebble isolation mass has no dependence on disc solid mass (see
equation \ref{eq:peb_iso}). In our simulations,
pebble isolation is not an issue unless the core migrates inside $\sim$0.1 AU;
what typically limits the core mass instead is the fact 
that the disc only has so much mass to give before it
drifts past the core.

\item {\em Pebble accretion depends sensitively on particle size and
    turbulent vertical stirring, and loses at least 1--2 orders of magnitude more
    mass to radial drift than is actually used to build cores.} Forming cores massive enough to nucleate gas giants
  (i.e., Earth-mass or larger objects) places stringent constraints on
  the parent disc.  Not only must the disc contain enough solids
  ($M_{\rm disc,s} > 30 M_\oplus$), but those solids should have
  aerodynamic stopping times not too short, and be embedded in gas
  that is not too turbulent. More quantitatively, over much of our
  parameter space, the final core mass increases exponentially 
  with $\tau/\alpha$ raised to some power
  (equations
  \ref{eq:mcore_final_tau_gt_alpha} and \ref{eq:mcore_final_tau_lt_alpha}),
  with
  $\tau \equiv \Omega t_{\rm stop} < 1$ the dimensionless measure of
  particle stopping time, $\Omega$ the orbital angular frequency, and
  $\alpha < 1$ the turbulent Mach number. As $\tau$ increases up to
  unity, the accretion cross-section of the core grows; as $\alpha$
  decreases, the vertical thickness of the solid disc decreases and
  the density of accreting particles increases. We find empirically
  that Jupiter-breeding cores require $\tau/\alpha > 0.1$.  
  Even when this condition is satisfied---and in this regard,
  the arguments by \citet{Pinte16}, \citet{Fung17},
  and \citet{Fung18} for low $\alpha$, practically
  inviscid discs are encouraging---pebble accretion is still
  wasteful in the sense that at least $\sim$90\%, and
  possibly much more of the solids can be lost
  to radial drift while growing a single core from a lunar mass to a few Earth masses 
   (see also \citealt{Ormel2017}, his section 7.1.4 and references therein). 
  In our models, the accretion efficiency of a single
  core may be $\sim$1--3\%
  (cf.~Figures \ref{fig:2x2_hist},
  \ref{fig:Mcore_v_t_compLJ14}, and \ref{fig:Mcore_v_t_compLJ14_lunar};
  see also \citealt{Guillot14, Lambrechts14,
  Ida16, Picogna18}). 
  This single-core
  efficiency is relevant for deciding whether a given disk
  has enough solids to generate even a single core
  massive enough to spawn a gas giant.

\item {\em Sub-Earths and gas giants, but no super-Earths---at least none
that avoid wholesale migration to the innermost disc edge.} Pebble
  accretion seems to be an all-or-nothing (and usually nothing)
  prospect: either sub-Earth cores remain sub-Earth, growing by less
  than a factor of 10 in mass, or conditions are tuned such  
  that cores grow rapidly while the disc is still gas-rich,
  leading to Jupiters. The exponential dependence of core mass on disc 
  properties (equation \ref{eq:mcore}) acts effectively as an on/off switch.
  When the switch is on, and cores grow to maximum, typically
  super-Earth size, they do not remain super-Earths, 
  but run away to become Jupiters under the early-stage,
  gas-rich conditions presumed by pebble accretion. 

 These results are robust against our assumption of a fixed
  particle size. More realistic grain growth models find that fragmentation
  of particle aggregates limits Stokes numbers $\tau \lesssim 0.1$ (larger
  $\tau$ leads to faster and more destructive particle collisions;
  e.g., \citealt{Birnstiel12}, their Figure 6).
  If we assume instead that particles
  have a fixed Stokes number $\tau \in \{0.01, 0.1\}$, 
  then we find the same general
  outcome: sub-Earths, Jupiters, and wholly migrated super-Earths,
  as seen in Figure \ref{fig:a0_scatter_fixtau}.
  Comparison with Figure \ref{fig:a0_scatter} reveals
  that this alternative assumption of constant $\tau$ produces
  a short-period pile-up of planets that extends to somewhat
  larger core masses (up to $\sim$10--20 $M_\oplus$),
  and discs that drain even
  faster (all discs have drift timescales $< 1$ Myr).
  
  Runaway can be avoided at ultra-small orbital distances where temperatures
  are high enough to stop gas accretion. Those super-Earths (1--10 $M_\oplus$)
  that do form in our models are all located at the innermost disc
  edge (0.01 AU), having migrated and piled up there. The problem
  is that short-period
  pile-ups of planets are not observed \citep{Lee17}; 
  nor is it clear how to disrupt the mean-motion resonant 
  chains that may result from such wholesale migration,
  in sufficient proportions to match observations
  (\citealt{Izidoro17}; but see \citealt{Goldreich14}).
  If migration were somehow suppressed,
  super-Earths could be formed at a variety of orbital distances,
  as we have verified by direct experimentation.
  Short of finding a mechanism to shut off migration (but see
  \citealt{Fung17} and \citealt{Fung18} for thoughts along these
  lines), we submit that 
  super-Earths are more naturally created later in a disc's 
  life, under gas-poor conditions where
  migration is not a concern, in a series of late-stage giant impacts
\citep[e.g.,][]{Lee14,dawson15,Dawson16,Inamdar16,Lee16,Ogihara18}.

  The gas giants that form by pebble
  accretion also generally undergo orbital migration. This is a simple
  consequence of the gas-rich conditions typically assumed by pebble
  accretion.
  The extent of migration exhibited by our model Jupiters
  ranges from shrinking the orbital radius by $\sim$30\%, to complete
  collapse from 10 AU to the innermost disc edge---and everything in between.
Had we adopted a more realistic (i.e., slower) prescription of gas accretion during the runaway phase (see \citealt{Machida10} and \citealt{Bitsch15b}), our model Jupiters would have migrated  farther, or been transformed into sub-Saturns---planets larger than Neptune but smaller than Saturn. 
  Contrary to the
  speculation that hot Jupiter cores can form in situ by pebble
  accretion \citep{batygin16}, we find that pebble isolation
  inside $\sim$0.1 AU limits core masses to values too low
  to trigger runaway gas accretion within the gas disc lifetime.
    
\item {\em Jupiters can form by pebble accretion, but in discs that
    may not fit mm-wave observations.} Pebble accretion favours
  particles large enough to have long stopping times $\tau$ (up to
  unity). The problem is that such particles are also the fastest to
  drain out of the disc. In nearly all of our Jupiter-producing runs,
  the disc is emptied of 0.01--1 cm sized pebbles on timescales
  ranging from 0.01--1 Myr. These drift times are troublingly shorter
  than the 1--10 Myr ages of discs seen in thermal mm-wave emission on
  scales of 10--100 AU. 
\end{enumerate}

Our findings, in particular our conclusions about the extreme sensitivity of pebble accretion rates to disc parameters, highlight
perhaps under-appreciated
difficulties in forming gas giants---and the apparent impossibility of forming super-Earths outside the innermost disc edge---by pebble accretion. For example, \citet[][LJ14]{Lambrechts14} reported that cores with initial masses of $10^{-3} M_\oplus$ readily grow by pebble accretion to 1--$10 M_\oplus$ within 1 Myr at orbital distances of 5--20 AU. Their result follows from assuming that (a) pebble accretion proceeds in the settling limit ($\tau \lesssim 1$), 
(b) the pebbles effectively comprise a 2D ultra-dense sheet whose scale height $H_{\rm s}$ is less than the accretion impact parameter $R_{\rm acc}$, and (c) Keplerian shear dominates the headwind in setting the relative velocity between pebble and core (cf.~our equation \ref{eq:vacc_settle}). Invoking these  assumptions gives the largest possible growth rates, with maximal accretion cross-sections and velocities set by Hill-sphere scales (their equation 28). 
In Figure \ref{fig:Mcore_v_t_compLJ14}, we compare
their optimistic solutions against other solutions
that do not make
the same assumptions. We focus on the sensitivity of the
evolution to $\alpha$, setting other parameters
such as the gas surface density, initial core mass,
and pebble size (see below for more comments
related to particle size) to the same values used
by LJ14. For $\alpha = 10^{-3}$, we see that accretion
rates can be overestimated by orders of magnitude.
Even if we assume that $\alpha = 0$ to strictly enforce
2D accretion, core growth can be delayed significantly,
as accretion begins and persists
in the hyperbolic regime for up to 1 Myr
at $a \sim 20$--40 AU.
Since such a delay is comparable to the gas disk lifetime, cores built by pebble accretion at large distances may not be able to nucleate gas giants, unless disk conditions are finely tuned or seed cores are more massive (cf.~Figure \ref{fig:Mcore_v_t_compLJ14_lunar}).

Note further that to 
ensure the comparison in Figure 
\ref{fig:Mcore_v_t_compLJ14} 
is fair, 
we incorporated LJ14's scheme for grain growth 
(their equations 20 and 25; see also \citealt{Birnstiel12}).
Their model disc is characterized by particle
stopping times $\tau \sim 0.01$--0.1
(their Figure 1); these are relatively high values that on the
one hand promote pebble accretion, but on the other lead to
fast radial drift and therefore disc sizes too compact compared
to mm-continuum observations. Figure \ref{fig:grain_growth} shows
that the LJ14 model of grain growth + drift (solid lines) does not
yield enough mm--cm objects at 20--60 AU to match
the size-wavelength relation exhibited by 
TW Hya, a 3--10 Myr system. 
Our models suffer from the same
problem, as we have noted under (iv) above.

\citet{Powell17} solved
the problem of reproducing the size-wavelength relation of
TW Hya by enhancing the disc's gas mass and slowing drift.
This seems to us 
an acceptable solution. Our comment would be that 
such a disc would not be conducive to pebble accretion
of mm--cm sized particles whose $\tau$'s would be too small
to accrete onto cores efficiently. Perhaps cores grow
instead from super-cm (larger $\tau$)
particles that are less visible to mm continuum observations.
Another way to explain the size-wavelength correlation
is to invoke optically thick substructures in the inner disc;
these could take the form of concentric rings of 
particles trapped in gas pressure maxima
(see \citealt{Tripathi18} for a specific discussion of UZ Tau,
and also our section \ref{ssec:future} below).

Our conclusions do not seem particularly sensitive to our
assumption of a single seed core. Were we to distribute seed cores
over a wide range of orbital distances, those close in would grow
more-or-less independently of those far out, since any given core 
diverts only a small fraction---a few percent at most---of the background
disc flow toward its own growth. If seed cores were packed so closely
as to be competing for the same solids, and if $R_{\rm acc} < H_{\rm s}$,
then growth would be ``neutral'' (section 8 of \citealt{Goldreich04}):
the distribution of relative core masses would not change, since
the doubling time
$M_{\rm core}/\dot{M}_{\rm core} \propto M_{\rm core}^0$
(equation \ref{eq:mdot_same}). If $R_{\rm acc} > H_{\rm s}$,
then growth would be ``orderly'': $M_{\rm core}/\dot{M}_{\rm core}$
would increase with increasing $M_{\rm core}$
(equations \ref{eq:mdot_hw_2d} and \ref{eq:mdot_shear_2d}),
and the distribution
of relative core masses would narrow \citep{Kretke14}.

\subsection{Future Directions}
\label{ssec:future}

Our model could use improvement in many respects.
There are, of course, the usual shortcomings resulting from our
incomplete understanding of orbital migration (e.g.,
the transition from Type I to Type II); runaway gas accretion; 
and perhaps most glaringly, disc turbulence and transport. 
We focus here on issues more specific to pebble accretion.

Revisiting the gas-particle dynamics of pebble accretion from first
principles seems worthwhile. Our paper is based 
on the equations of \citet{Ormel10}, which
assume a strict Cartesian shear for the background gas disc.
But cores perturb gas streamlines onto horseshoe orbits in 2D
\citep[e.g.,][their Figure 12]{Ormel13} and ``transient horseshoes''
in 3D \citep[e.g.,][]{fung15}, either of which can deflect particles,
particularly those with small $\tau$,
away from the core, and conceivably radically altering accretion
probabilities. \citet{Xu17} have tested some of the scaling relations
of \citet{Ormel10} using 3D simulations, but only under
restrictive conditions and with mixed results. 
A more comprehensive
study, starting with re-deriving accretion cross sections and
velocities for 2D laminar flow patterns, would be welcome (see also
\citealt{popovas18}).

Images of protoplanetary discs from the Atacama Large Millimeter
Array have revealed concentric rings of dust (\citealt{ALMA15};
\citealt{Andrews16}; S.~Andrews 2017, personal communication).
These rings may trace local gas pressure maxima that can ``trap''
inflowing pebbles (see the review by \citealt{Pinilla2017}
and references therein). Are these rings the sites of planetesimal/planet
formation? How does pebble accretion proceed in the
presence of traps? In particular, how would the accretion
rate (\ref{eq:mdot_same}) and the efficiency (\ref{eq:epsilon})
change if the seed core were located at the centre of the pressure bump?
Because drift speeds slow to zero in traps, interparticle collisions
are gentler, and bodies may stick their way to larger sizes 
and larger Stokes numbers
(cf.~\citealt{Chiang10}, their section 4),
allowing for more rapid pebble accretion. 
How long the solids in the trap
take to congeal into a single body is an outstanding question.

\section*{Acknowledgments}
We thank Bertram Bitsch, Jeffrey Fung, Anders Johansen, Michiel Lambrechts, Wladimir Lyra, Ruth Murray-Clay, Chris Ormel, Diana Powell, Michael Rosenthal, Julia Venturini, and an anonymous referee for comments that helped to improve the manuscript. 
JWL and EC acknowledge support from the National Science Foundation and a Berkeley BEAR grant.
EJL is supported by the Sherman Fairchild Fellowship from Caltech. This research used the Savio computational cluster resource provided by the Berkeley Research Computing program at the University of California, Berkeley (supported by the UC Berkeley Chancellor, Vice Chancellor for Research, and Chief Information Officer). 

\appendix

\section{Regimes of Core Growth by Pebble Accretion}
\label{app:eff}

\begin{figure}
    \centering
    \includegraphics[width=0.5\textwidth]{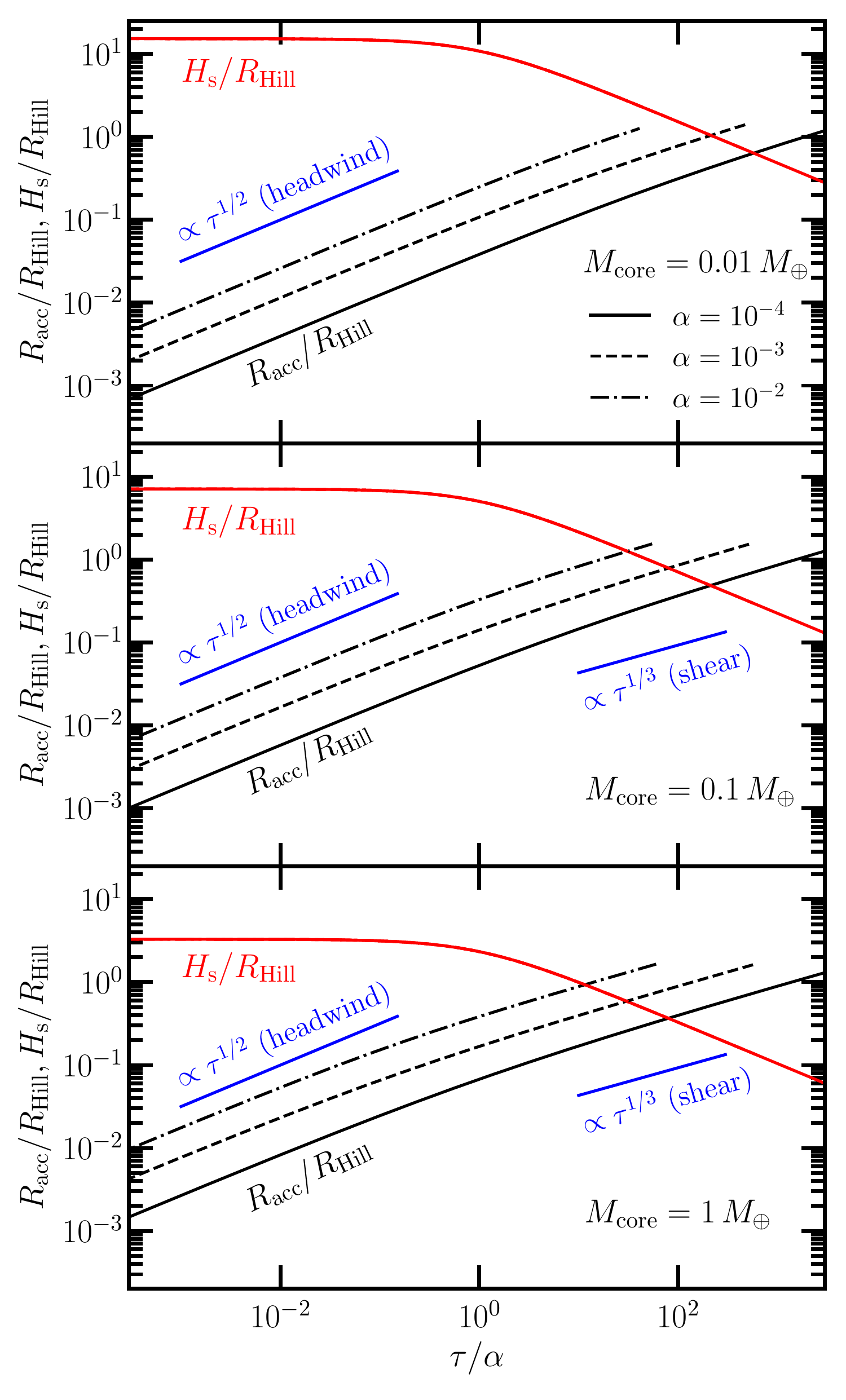}
    \vspace{-5mm}
        \caption{Comparison between the accretion radius $R_{\rm acc}$
      (black lines) and the particle scale height $H_{\rm s}$ (red
      lines) as a function of the ratio between particle Stokes number
      (dimensionless stopping time) $\tau$ and turbulent Mach number
      $\alpha$. We truncate an $R_{\rm acc}/R_{\rm Hill}$ curve when pebble
      accretion ceases to be in the settling regime (i.e., when
      equation \ref{eq:settle_condition} is not satisfied). All values
      are calculated at $a = 5$ AU. At fixed core mass, accretion is headwind-dominated at small $\tau/\alpha$ and shear-dominated at large $\tau/\alpha$ (compare $R_{\rm acc}/R_{\rm Hill}$ curves to blue segments; see equations
      \ref{eq:Racc_hw} and \ref{eq:sheardom}).
}
    \label{fig:accr_rad_Hs}
\end{figure}

Over much of our parameter space, the accretion radius
$R_{\rm acc}$ is smaller than the particle scale height $H_{\rm s}$. 
For a small subset of our models, at core masses above a few $M_\oplus$,
this inequality reverses
(see Figure \ref{fig:accr_rad_Hs}; to be clear, the equations
we solve are general enough to accommodate this possibility).
We discuss here how the final core mass changes its scaling behaviour
with time and other variables across parameter space (cf.~equations 
\ref{eq:mdot_same}--\ref{eq:mcore_final_tau_lt_alpha}), 
assuming throughout that pebble
accretion is in the settling regime (stopping time $\tau \lesssim 1$).
See section \ref{sssec:settling} for more background.

When the headwind parameter $\zeta = v_{\rm hw}/v_{\rm Hill}$ exceeds
the accretion radius $b = R_{\rm acc}/R_{\rm Hill}$,
the accretion velocity is dominated by the headwind
velocity: $v_{\rm acc} \sim v_{\rm hw}$, which for our disc temperature
profile is constant. The accretion radius
$R_{\rm acc}$ follows from equating the kick velocity $\Delta v$
(equation \ref{eq:kick}) to $v_{\rm acc}/4$ (OK10):
\begin{align}
    \frac{GM_{\rm core}}{R_{\rm acc,hw}^2}\frac{\tau}{\Omega} &\sim \frac{v_{\rm hw}}{4} \nonumber \\
    R_{\rm acc,hw} &\sim \left(\frac{4GM_{\rm core}\tau}{\Omega v_{\rm hw}}\right)^{1/2}.
    \label{eq:Racc_hw}
\end{align}
This is equivalent to the wind-shearing radius
of \citet{perets11}. Under headwind-dominated conditions,
$R_{\rm acc,hw}$ exceeds the particle scale height
$H_{\rm s} = H\sqrt{\alpha/(\alpha+\tau)}$
when
\begin{align}
\frac{\tau}{\alpha} &> \left(\frac{f_P}{4\alpha}\right)^{1/2}\left(\frac{c_{\rm s}}{\Omega a}\right)^{2}\left(\frac{M_\star}{M_{\rm core}}\right)^{1/2} \nonumber \\
&\gtrsim 100 \left(\frac{a}{1\,{\rm AU}}\right)^{1/2}\left(\frac{0.01M_\oplus}{M_{\rm core}}\right)^{1/2}\left(\frac{0.001}{\alpha}\right)^{1/2}
\label{eq:tau_alpha_hw}
\end{align}
where the second inequality follows from our disc parameters: $f_P
= 11/8$, $T = 260\,{\rm K}(a/1\,{\rm AU})^{-1/2}$, and $M_\star =
M_\odot$. Technically the above criterion for whether $R_{\rm acc}$
exceeds $H_{\rm s}$ is derived assuming $\tau >\alpha$,
but this is a safe assumption; if $\tau < \alpha$, so
that small particles are strongly stirred by gas turbulence, then
only super-massive cores ($\gtrsim 100 M_\oplus$) have accretion
radii $R_{\rm acc} > H_{\rm s}$. 

On the other hand, when $\zeta < b$, the accretion velocity is dominated
by the Keplerian shearing velocity
$v_{\rm acc} \sim 3\Omega R_{\rm acc}/2$. Then
\begin{align}
    \frac{GM_{\rm core}}{R_{\rm acc,sh}^2}\frac{\tau}{\Omega} &\sim \frac{3\Omega R_{\rm acc,sh}}{8} \nonumber \\
    R_{\rm acc,sh} &\sim \left(\frac{8GM_{\rm core}\tau}{3\Omega^2}\right)^{1/3}. \label{eq:sheardom}
\end{align}
The transition from headwind-dominated to shear-dominated accretion
(what \citealt{Lambrechts12} call drift accretion vs.~Hill accretion)
occurs when $R_{\rm acc,hw} \sim R_{\rm acc,sh}$:
\begin{align}
M_{\rm core,trans} &\sim v_{\rm hw}^3/(9G\Omega\tau) \nonumber\\
    &\sim 0.1\,M_\oplus\left(\frac{0.001}{\tau}\right)\left(\frac{a}{1\,{\rm AU}}\right)^{3/2}.
\end{align}
We emphasize that the transition core mass depends on $\tau$
(see also equation 7.9 of \citealt{Ormel2017}).
Under shear-dominated conditions, $R_{\rm acc} > H_{\rm s}$ when
\begin{align}
\frac{\tau}{\alpha} &> \left(\frac{3}{8\alpha}\right)^{2/5}\left(\frac{c_{\rm s}}{\Omega a}\right)^{6/5}\left(\frac{M_\star}{M_{\rm core}}\right)^{2/5} \nonumber \\
&\gtrsim 30 \left(\frac{a}{1\,{\rm AU}}\right)^{3/10}\left(\frac{M_\oplus}{M_{\rm core}}\right)^{2/5}\left(\frac{0.001}{\alpha}\right)^{2/5}
\end{align}
where we have again safely assumed $\tau > \alpha$.

Armed with the above relations,
we derive the approximate scaling behaviour of the 
core mass in various regimes. When $R_{\rm acc} < H_{\rm s}$, 
the accretion rate $\dot{M}_{\rm core}$ is identical 
between headwind and shear-dominated regimes,
and is given by equation (\ref{eq:mdot_same}).
The accretion rate is the same between these regimes
because $\dot{M}_{\rm core} \propto R_{\rm acc}^2 v_{\rm acc}$
(when $R_{\rm acc} < H_{\rm s}$), and 
the combination $R_{\rm acc}^2 v_{\rm acc}$ is given in the settling
limit by its defining condition, 
$GM_{\rm core}/R_{\rm acc}^2 \times \tau/\Omega \sim v_{\rm acc}/4$ \citep{Ormel10}.
The end result is that $M_{\rm core}(t)$
is given by equation (\ref{eq:mcore}) and is exponential if not
super-exponential in time, depending on how $\Sigma_{\rm s}$
evolves at the position of the core.

When $R_{\rm acc} > H_{\rm s}$, and when accretion is headwind-dominated,
$\dot{M}_{\rm core} = 2\Sigma_{\rm s}R_{\rm acc,hw}v_{\rm hw}$, which together
with (\ref{eq:Racc_hw}) yields
\begin{equation}
   \label{eq:mdot_hw_2d}
   \dot{M}_{\rm core} = 4\left(\frac{v_{\rm hw}}{\Omega a}\right)^{1/2}\tau^{1/2}\left(\frac{M_{\rm core}}{M_\star}\right)^{1/2}\Omega\Sigma_{\rm s}a^2
\end{equation}
and
\begin{equation} 
\label{eq:mcore_hw_2d}
M_{\rm core}(t) = \left[M_{\rm core}(0)^{1/2} + 2\left(\frac{v_{\rm hw}}{\Omega a}\right)^{1/2}\tau^{1/2}\frac{\Omega a^2}{M_\star^{1/2}} \int_0^t \Sigma_{\rm s} dt \right]^2 
\end{equation}
where all quantities are evaluated at the core's position.
When $R_{\rm acc} > H_{\rm s}$ and accretion is shear-dominated,
$\dot{M}_{\rm core} = 3\Sigma_{\rm s}R_{\rm acc,sh}^2\Omega$,
so that
\begin{equation}
    \label{eq:mdot_shear_2d}
    \dot{M}_{\rm core} = (192)^{1/3}\tau^{2/3}\left(\frac{M_{\rm core}}{M_\star}\right)^{2/3}\Omega\Sigma_{\rm s}a^2
\end{equation}
and
\begin{equation} \label{eq:mcore_shear_2d}
M_{\rm core}(t) = \left[M_{\rm core}(0)^{1/3} + \left(\frac{8}{3}\right)^{2/3} \tau^{2/3} \frac{\Omega a^2}{M_\star^{2/3}} \int_0^t \Sigma_{\rm s}dt\right]^3 \,.
\end{equation}

\bibliographystyle{mnras}
\bibliography{references}

\bsp	
\label{lastpage}
\end{document}